\documentclass{emulateapj}

\newcommand{\Msun}{\mbox{M$_{\odot}$}}
\newcommand{\Rsun}{\mbox{R$_{\odot}$}}

\newcommand{\kms} {\mbox{km~s$^{-1}$}}

\newcommand{\ltsimeq}{\raisebox{-0.6ex}{$\,\stackrel
        {\raisebox{-.2ex}{$\textstyle <$}}{\sim}\,$}}
\newcommand{\gtsimeq}{\raisebox{-0.6ex}{$\,\stackrel
        {\raisebox{-.2ex}{$\textstyle >$}}{\sim}\,$}}
\def\lesssim{\mathrel{\hbox{\rlap{\hbox{\lower4pt\hbox{$\sim$}}}\hbox{$<$}}}}
\def\gtrsim{\mathrel{\hbox{\rlap{\hbox{\lower4pt\hbox{$\sim$}}}\hbox{$>$}}}}
\def\ggg{\mathrel{\hbox{\rlap{\hbox{\lower4pt\hbox{$\sim$}}}\hbox{$>$}}}}

\slugcomment{ }

\shorttitle{ KOI-54: Tidally-Excited Pulsations and Brightenings}
\shortauthors{Welsh et al.}

\begin{document}

\title{ KOI-54: The {\it Kepler} Discovery of Tidally-Excited Pulsations 
and Brightenings in a Highly Eccentric Binary}

\author{
William F.~Welsh$^1$,
Jerome A.~Orosz$^1$,
Conny Aerts$^{2}$,
Timothy M.~Brown$^{3,17}$, 
Erik Brugamyer$^{4}$,
William D.~Cochran$^{4}$,
Ronald L.~Gilliland$^{5}$,
Joyce Ann Guzik$^{6}$,
D.~W.~Kurtz$^{7}$,
David W.~Latham$^{8}$,
Geoffrey W. Marcy$^{9}$,
Samuel N.~Quinn$^{8}$,
Wolfgang Zima$^{2}$,
Christopher Allen$^{16}$,
Natalie M.~Batalha$^{10,11}$, 
Steve Bryson$^{11}$,
Lars A.~Buchhave$^{14,19}$,
Douglas A.~Caldwell$^{11,12}$, 
Thomas N.~Gautier III$^{13}$,
Steve B.~Howell$^{11}$,
K. Kinemuchi$^{11,15}$,
Khadeejah A.~Ibrahim$^{16}$,
Howard Isaacson$^{9}$,
Jon M.~Jenkins$^{11,12}$,
Andrej Prsa$^{18}$,
Martin Still$^{11,15}$,
Rachel Street$^{3}$,
Bill Wohler$^{16}$,
David G.~Koch$^{11}$, and 
William J.~Borucki$^{11}$
}

\affil{$^1$ Astronomy Department, San Diego State University,
  San Diego, CA 92182 USA}
\affil{$^{2}$ Instituut voor Sterrenkunde, K.U.Leuven, Celestijnenlaan 
 200D, B-3001 Leuven, Belgium; and Department of Astrophysics, IMAPP, 
 University of Nijmegen, PO Box 9010, 6500 GL Nijmegen, The Netherlands}
\affil{$^{3}$ Las Cumbres Observatory Global Telescope, 6740 Cortona Dr. 
 Ste. 102, Goleta, CA 93117 USA}
\affil{$^{4}$ McDonald Observatory and Department of Astronomy, 
 The University of Texas at Austin, Austin, TX 78712 USA}
\affil{$^5$ Space Telescope Science Institute, Baltimore, MD 21218}
\affil{$^{6}$ XTD-2, MS T086, Los Alamos National Laboratory, Los Alamos, 
 NM 87545-2345 USA}
\affil{$^7$ Jeremiah Horrocks Institute of Astrophysics,
University of Central Lancashire, Preston PR1 2HE, UK}
\affil{$^8$ Harvard-Smithsonian Center for Astrophysics, 60 Garden Street, 
 Cambridge, MA 02138 USA}
\affil{$^{9}$ Department of Astronomy, University of California, Berkeley, 
CA 94720 USA}
\affil{$^{10}$ Department of Physics and Astronomy, San Jose State 
University, San Jose, CA 95192 USA}
\affil{$^{11}$ NASA Ames Research Center M/S 244-30, Moffett Field, CA 
 94035 USA}
\affil{$^{12}$ SETI Institute, Mountain View, CA 94043 USA}
\affil{$^{13}$ Jet Propulsion Laboratory/California Institute of 
 Technology, Pasadena, CA 91109 USA}
\affil{$^{14}$ Niels Bohr Institute, University of Copenhagen,
DK-2100 Copenhagen, Denmark}
\affil{$^{15}$ Bay Area Environmental Research Inst., Inc., 560 Third St. 
 W, Sonoma, CA 95476 USA}
\affil{$^{16}$ Orbital Sciences Corporation/NASA Ames Research Center, 
 Moffett Field, CA 94035 USA}
\affil{$^{17}$ Department of Physics, University of California, Santa 
 Barbara, CA 93106 USA}
\affil{$^{18}$ Dept. of Astronomy and Astrophysics, Villanova University, 
800 E. Lancaster Ave., Villanova, PA 19085 USA}
\affil{$^{19}$Centre for Star and Planet Formation, Natural History 
Museum of Denmark, Univer.\ Copenhagen, DK-1350 Copenhagen, Denmark}

\altaffiltext{$\dagger$}{Based in part on observations obtained at the 
W.~M.~Keck Observatory, which is operated by the University of 
California and the California Institute of Technology.}

\altaffiltext{*}{To whom correspondence should be addressed:\\  
E-mail: wfw@sciences.sdsu.edu}

\begin{abstract}
{\it Kepler} observations of the star HD~187091 (KIC~8112039, 
hereafter KOI-54)  revealed 
a remarkable light curve exhibiting sharp periodic brightening 
events every 41.8 days with a superimposed set of oscillations forming a 
beating pattern in phase with the brightenings. Spectroscopic 
observations revealed that this is a binary star with a highly eccentric 
orbit, e=0.83. We are able to match the {\it Kepler} light curve and 
radial velocities with a nearly face-on ($i$=5.5 degree) 
binary star model in which the 
brightening events are caused by tidal distortion and irradiation of 
nearly identical A stars during their close periastron passage.  The 
two dominant oscillations in the light curve, responsible for the 
beating pattern, have frequencies that are the 91st and 90th harmonic 
of the orbital frequency. 
The power spectrum of the light curve, after 
removing the binary star brightening component, reveals a large number 
of pulsations, 30 of which have a signal-to-noise ratio $\gtsimeq$7.
Nearly all of these pulsations have frequencies that are either integer 
multiples of the orbital frequency or are tidally-split multiples of 
the orbital frequency.
This pattern of frequencies unambiguously establishes the pulsations 
as resonances between the dynamic tides at periastron and the free
oscillation modes of one or both of the stars. KOI-54 is only the 4th 
star to show such a phenomenon, and is by far the richest in terms of 
excited modes. 

\end{abstract}

..........................................................

\section{Introduction}
In its search for transiting extrasolar planets, the {\it Kepler} Mission
obtains high-precision time series photometry of $\sim$156,000 stars 
(Koch 2010, Borucki 2010). 
This very large high-signal-to-noise sample, combined with the 
unprecedented near-continuous long-timescale coverage, results in a 
high potential for discovery of astrophysically important stars.
KOI-54 is one of those serendipitous discoveries.

Cataloged as HD~187091 before being flagged as {\it Kepler Object 
of Interest} 54, this star was  classified as a single bright (V=8.38) 
A star. The object did not seem unusual in any way prior to these 
{\it Kepler} observations and  attracted little attention. 
This is in stark contrast to the astonishing {\it Kepler} light curve, 
which immediately offered a challenging puzzle begging for elucidation 
--- see Fig.~1. 

In this paper, we present the {\it Kepler} observations and a large
set of precise radial velocities that led to our binary star 
model for KOI-54. We investigate the rich set of oscillations that 
are present, and conclude that these are tidally-driven pulsations.
We also explore the evolutionary status of these stars and find a 
self-consistent scenario for the binary system.
We finish with some remarks on the puzzles that still remain, and
discuss in some detail the systematic issues that could potentially 
affect our interpretation, but find these issues to generally have 
no significant effect on our conclusions.

\section{Observations}

\subsection{Overview of the Light Curve}

KOI-54 has been observed by {\it Kepler} almost continuously from
2009 May 02 during Commissioning observations (Q0) though 
2010 Mar 22 ``Quarter 4'' (Q4). The light curve exhibits two 
remarkable features: a periodic brightening
spike of $\sim$0.7\% occurring every 42 days and a $\sim$0.1\%  
``beat pattern'' of pulsations in phase with the brightening events.
The beat pattern arises from the interference of two pulsations with 
periods near 11 hours. The pulsations continue during the brightening
and repeat from event to event -- they are strictly in phase with the 
brightening.

An initial hypothesis that the repeating brightening events
might follow from near-field microlensing (Sahu and Gilliland 2003)
was pursued.  Such a large event would result from a black hole
of several solar masses orbiting and transiting an A star.
Motivated by this hypothesis, a 1958~s duration {\it SWIFT} \ XRT 
observation was made on 2010 Apr 25, during one of the brightening 
events. The source was not detected in the 0.3--10 keV bandpass, 
but we estimate 
the 5-sigma upper limit on the X-ray flux to be $F < 1.23 \times 
10^{-13}$ $\mathrm{ erg \ cm^{-2} \ s^{-1} }$. Using the {\it HIPPARCOS} 
parallax of 3.14 milliarcsec, 
the upper limit to the luminosity is
 $\rm L < 1.5 \times 10^{30}$ erg s$^{-1}$. 
Given the null detection, we make no further discussion of these 
observations.

Radial velocity observations quickly ruled out the transiting black 
hole hypothesis because no large rapidly changing reflex motion of 
the A star was observed. Also, no ready explanation of the oscillations 
existed in this scenario, and when examined in detail the 42-day 
brightenings deviate 
from the expected shape of microlensing events.
Instead, as we show below, KOI-54 is a highly eccentric binary 
star system, and the brightening events are caused by tidal distortion 
and irradiation of the two stars during their close periastron 
passage. The two dominant pulsations producing the beating seen
in the light curve have frequencies that are exactly 90 and 91 times 
the orbital frequency, hence result from tidally driven pulsation modes.

\subsection{The {\it{Kepler}} Photometry and Calibration}
The {\it{Kepler}} CCDs are read out every 6.54~s (6.02~s live-time) 
and co-added on-board. In Short Cadence (SC) mode the signal is 
combined to achieve approximately 58.8~s sampling cadence, and in 
Long Cadence (LC) mode the data are binned to 29.424~min cadence.
KOI-54 is heavily saturated in these 6~s exposures, and blooming
affects approximately 30 pixels of the 74-pixel aperture 
(approximately 7 pixels diameter plus two columns of 25 pixels).
However, the electrons are not lost, and because of {\it{Kepler's}} 
stability, superb relative photometry is achievable ---
see \citet{Gilliland10} for a discussion.
For more details on the design and performance of the {\it{Kepler}} 
photometer see \citet{Koch10} and Jenkins et al.~(2010a,b).

Each {\it{Kepler}} pixel spans $\sim$4 arcseconds so the very large 
photometric aperture includes a few very faint (by comparison)
background stars. 
The brightest star (KIC~8112007) has Kp=17.6 (4850 times fainter 
than KOI-54), and ground-based images with 1 arcsecond seeing
show no star brighter than 18th magnitude within 5 arcseconds.
We conclude that background starlight contaminates the light curve 
by less than 250 ppm (0.025\%), and thus variations in the light 
curve are intrinsic to KOI-54.

In addition to gaps due to the scheduled quarterly spacecraft rolls,
safe-mode events and other spacecraft anomalies are present; 
cosmic rays and other noise sources also contaminate the signal.
A simple automated sigma--threshold rejection method could not be used
because of the presence of the complex oscillation signal, so the light
curve was carefully examined and obvious outliers were omitted by hand:
158 points were removed, leaving a total of 14,277 observations.
This corresponds to a duty cycle of an astounding 92.5\% over the 
span of 10.6 months.

The observations span 321.7~d, nearly eight complete 41.8~d orbital 
cycles. For each spacecraft roll, the target is on a different CCD
with pixels of different sensitivity, and
thus jumps in the light curve occur from  Quarter to Quarter.
Mean fluxes over the 4 Quarters span 1.01 -- 1.07 $\times 10^{10}$ 
e$^{-}$/cadence. To correct for these changes, and the more troublesome 
safemode-related, pointing-adjustment, and other medium-timescale 
discontinuities, the ``RAW'' pipeline processed light 
curves\footnote{Data are available at the Multimission Archive (MAST) 
at the Space Telescope Science Institute:
{\url{http://archive.stsci.edu/kepler/data\_search/search.php}}}
were detrended in the following way.
First, the brightening events were masked in orbital phase from 0.9 to 
1.1 (i.e., 20\% of the light curve). Then using gaps in the time series 
to define sections, the sections were fit with a low-order polynomial 
(typically cubic, though linear or a constant was used if appropriate). 
The polynomial was then subtracted from the time series, and the 
remainder divided by the mean of the polynomial. A value of +1.00 was 
then added to this quotient giving a relative flux with a mean of unity. 
The pipeline uncertainty estimates were boosted by a factor of 6,
based on the rms scatter of the residuals of an early model fit.
This scaling factor was chosen for simplicity, though each Quarter
should have its own scaling (e.g. values of 5.0, 5.5 and 7.3 for three 
different Quarters were found). The scatter in the residuals was 
significantly in excess of what was expected just from the error bars, 
but in hindsight this was due to using too few sinusoidal
components in our initial modeling (see below).
Slightly overestimating the uncertainties
has no adverse affect on the estimation of the stellar and 
orbital parameters, and helps account for systematic noise 
(e.g. photometric trends after a safe mode or other anomaly) 
that is not included in the statistical uncertainties.

One month of Short Cadence data was obtained during Quarter 3;
for the characteristics of {\it Kepler} SC data, see
\citet{Gilliland10}.
These observations contain 43,990 points and span 30.035 days. 
Examination of these SC data, and their power spectrum at
higher frequencies, did not reveal any features not already
well resolved in the LC data, therefore only the Long Cadence
data are used for our analyses.

Finally, the relative timing precision is good to 0.5~s, but the 
absolute timing is uncertain by as much as 6.5~s, and this systematic 
uncertainty should be added to the statistical uncertainty in the 
epoch of periastron $T_{p}$ in Table~2 --- see {\it Kepler Data 
Release~8 Notes} \citep{Machalek10}.

\subsection{Spectroscopy and Radial Velocities}

A variety of telescopes and instruments were used to provide moderate 
to high resolution spectroscopy for radial velocity (RV) and spectral 
modeling. The spectra show double sets of absorption lines revealing 
the binary star nature of the system, and in Fig.~2 we show a 
section of a Keck HIRES spectrum where pairs of lines are clearly 
exhibited. The stars are of similar spectral type and luminosity.
To determine the stellar parameters we carried out an analysis of 
11 relatively clean Fe~I and Fe~II lines using the LTE spectral 
synthesis code {\sc MOOG} \citep{sneden73}, modified for binary 
star analysis. Using an input linelist and two separate model 
atmospheres, the code computes a synthetic spectrum for each 
stellar component. Given a velocity separation and luminosity ratio, 
the code then overplots the resulting spectra onto the observed spectrum. 
We then varied the stellar parameters (temperature, gravity, metallicity 
and microturbulent velocity) until the best overall match to our selected 
iron lines was found. 
Using an HET HRS spectrum taken very near 
periastron (when the stars were cleanly separated in Doppler velocity 
by 25.5 \kms), we obtained the following information about the stars. 
The stellar temperatures T$_{1}$ and T$_{2}$, are 8500 and 8800 K 
$\pm$200 K, with log g of 3.8 and 4.1 $\pm$0.2, and a luminosity 
ratio L$_{2}$/L$_{1}$ of 1.22 $\pm$0.04 for the wavelength range covered 
in the spectral analysis, $\sim$4500--6500 \AA. 
Both stars are $\sim$ 2--3 
times more metal-rich than solar with $\rm [Fe/H]$ = 0.4 $\pm$0.2. 
Thus the stars are both A-type near-main sequence stars, and potentially 
pulsating $\delta$ Scuti variables --- see \citet{Aerts10} for a 
thorough discussion of $\delta$ Scuti and other pulsating stars. 

Estimates of $V_{\rm rot} \sin{i}$  depended on the spectra and method 
used, but all indicate a low projected velocity ranging from $\ltsimeq$5 
to 10 \kms. Given the difficulty in measuring $V_{\rm rot} \sin{i}$ we
adopt a value of 7.5 $\pm$4.5 \kms.
When corrected for the very low inclination of the system 
(discussed in \S3) the true $V_{rot}$ can be $\ltsimeq$ 50-100 \kms, and 
this is still relatively low for an A-star. This low rotation rate, 
combined with the metal-rich abundance, suggests these are chemically 
peculiar Am or possibly Ap stars.
An inspection of the Nd~III 6145\AA~ and PrIII~6160\AA~ lines show that 
they are not strong, as they usually are in roAp stars.

A total of 51 pairs of radial velocities were obtained in 2010
from six different telescopes+spectrographs, and these are listed in 
Table~1.
To minimize any potential systematic offset between the velocities 
acquired with different instruments, all RV observations were calibrated 
using the same procedure and all velocities measured
using the TODCOR technique \citep{Zucker94}. Every velocity 
measurement is referenced to a common radial velocity standard, 
HD~182488, that was observed every night, following standard practice for 
{\it Kepler} follow-up observations of targets of interest. 
Although not simultaneous with the {\it Kepler} photometry presented 
in this paper, these observations provided the key to understanding 
the nature of KOI-54: the stars are on a highly eccentric orbit,
e=0.83, with periastron passage at the times of the brightenings. 
Thus the mutual interaction of the stars when closest together 
produces the brightening events seen in the {\it Kepler} photometry, 
and this is discussed in detail in \S 3 and \S 5.1.
Fig.~3 shows the radial velocities and Keplerian fits, and Table 2 
includes the orbital elements from the radial velocity-only fit.
The ratio of K-velocities gives a mass ratio of 1.034, again 
confirming the similarity of the two stars. Note: The TODCOR
methodology we employed for measuring the radial velocities does not 
provide uncertainty estimates on the radial velocities, so the mean 
of the residuals from the radial velocity fit to the observations was
assigned as the uncertainty to all the velocities: 0.31 \kms. Given the 
relatively good fit to the RV data over most of the orbit, this 
approximation is justifiable. Concerns over systematic errors, as
seen in the residuals of the fit, are discussed in \S5.1.3.

\section{Modeling}
We employ a modified version of the {\sc ELC} modeling code of 
Orosz \citep{Orosz00} to simultaneously fit both the photometry and 
radial velocities.
The non-spherical stars are covered with a fine grid of tiles, and 
for each time the intensity and velocity of the tiles are summed 
to produce the light curve and radial velocities. 
Gravitational distortions are modeled assuming a standard Roche 
potential, including the rotation of the stars themselves; the
potentials are recomputed at each orbital phase because of the 
elliptical orbit
--- see the appendix of \citet{Orosz00} for a description of the 
potential that is based on \citet{Avni75}, and see
\citet{Sepinksy07} for a thorough discussion of equipotential surfaces 
in nonsynchronous eccentric binaries.
Gravity darkening is included, using an exponent $\beta$=0.25 
appropriate for early-type radiative stars \citep{vonzeipel24, claret00}.
We used tabulated spherical {\sc NextGen}/PHOENIX model intensities 
(Hauschildt et al.~1997). The model flux is then integrated over the 
{\it Kepler} spectral response (approximately 4250--8950~\AA, peaking 
at 5890~\AA \ with a mean wavelength of 6400~\AA \ --- see 
{\citet{VanCleve09} and \citet{Koch10}). 
Irradiation of the stars is handled following the standard 
prescription of \citet{Wilson90}:
The light from each star is the sum of (1) the intrinsic intensities
of each tile (modified for the local gravity, and if blackbodies 
are used, the limb darkening); and (2) an irradiation ``reflection'' 
component on the inward-facing hemispheres.
The irradiation modifies the local temperature in the following way:
$T^{\prime}{}^{4} = T^{4} \times
\left[ 1 + A_{bol} \frac{F_{irr}}{F} \right]$
where $T$ and $F$ are the temperature and bolometric flux of the star,
and $F_{irr}$ is the incident bolometric flux from the companion star.
$A_{bol}$ is the bolometric albedo (not to be confused with the Bond 
albedo) and is the ratio of re-radiated to incident energy.
A radiative atmosphere has $A_{bol}$=1 (implying local energy
conservation), and we hold $A_{bol}$=1 in our models.
\citet{Kallrath99} give an excellent description of Wilson's method
and we refer the reader to that source for more details.
At periastron, the maximum change in temperature over the surface of the 
stars, including gravity darkening and irradiation, is 88~K (=1\%) 
for Star 1 and 61~K (0.7\%) for Star 2. At apastron, the difference is 
only 0.5 and 0.4 K.

The free parameters in the model are:
the stellar masses M$_{1}$ and M$_{2}$, radii R$_{1}$ and R$_{2}$, 
temperatures T$_{1}$ and T$_{2}$,
and 6 orbital parameters: inclination $i$, orbital period $P$, 
epoch of periastron $T_{p}$, argument of periastron $\omega$, 
eccentricity $e$, and systemic velocity $\gamma$ (held the same 
for both stars).
The temperature is essentially unconstrained by the single-color 
broadband photometry, but the temperature ratio is weakly constrained
by the ratios of radii and luminosities.
While the models had freedom to vary the temperatures, the solution 
remained at the input spectroscopic temperatures.
The individual radial velocity measurements are fit, not just the 
$K_{1}$ and $K_{2}$ radial velocity amplitudes from the RV-only fit
(i.e.\ we do not adopt the RV-only solution).
The rotation of the stars, usually defined via the ratio 
$\Omega = \Omega_{rot}/\Omega_{orb}$, where $\Omega_{rot}$ is the 
stellar rotation angular frequency and $\Omega_{orb}$ is the orbital 
angular frequency, were also treated as free parameters.
But for KOI-54  we defined these as 
$\Omega \equiv \Omega_{rot}/\Omega_{ps}$ where $\Omega_{ps}$ is the
``pseudosynchronous'' rotation frequency and depends only on the 
eccentricity and orbital period as given by eqn.\ (42) in Hut (1981):}
\begin{equation}
\Omega_{ps} = (2 \pi/ P) \times 
\frac{ 1 + \frac{15}{2}e^{2} + \frac{45}{8}e^{4} + \frac{5}{16}e^{6} }
     { ( 1 + 3e^{2} + \frac{3}{8}e^{4} )  ( 1 - e^2 )^{3/2}  }
\end{equation}
For an elliptical orbit true synchronous rotation is impossible, 
but there is a pseudosynchronous spin such that over the course of 
an orbit there is no {\em net} torque on the star's rotation,
and so the spin will not evolve: $\dot{\Omega}=0$  (Hut 1981). 
Note that $\Omega_{ps}$ is $\sim$20\% slower than the orbital frequency 
at periastron, i.e., the spin is slower than what is necessary to keep 
the star tidally locked at periastron passage. 

In addition, other ``observed parameters'' are used to 
constrain the model: $V_{rot}\sin{i}$, log g, and the ratio of 
luminosities L2/L1.
These are not fixed in {\sc ELC}, rather, the models are steered toward 
them via a $\chi^{2}$ penalty. A genetic algorithm and Markov chain 
Monte Carlo are used to find the best-fit models (in a $\chi^{2}$ sense) 
and confidence intervals. The best-fit model has a reduced $\chi^{2}$ 
less than 1, indicating the uncertainties in the light curve were 
overestimated in the data calibration; 
but we determined the 1-$\sigma$ parameter intervals in the 
standard way by marginalizing over all other parameters and determining 
the interval bounded by $\chi^{2}_{min} + 1$; we did not decrease this
interval to account for the reduced $\chi^{2}$ being less than 1.

\subsection{ELCsinus}
The {\sc ELC} code models the binary star light curve and radial velocity, 
but does not model the oscillations. Attempts to isolate the binary light 
curve from the pulsation light curve proved inadequate for the
high precision {\it Kepler} observations, so we added the following 
functionality into the {\sc ELC} code, now dubbed ``{\sc ELCsinus}''.
First, a trial binary star light curve is subtracted from the observed
light curve. The mean is then subtracted and the Fourier transform 
is computed. The 15 largest peaks in the power spectrum are found
(omitting sidelobes due to leakage of the 2 dominant peaks), 
and a sum of sines 
and cosines is made using the 15 measured amplitudes and phases. This 
15-sine pulsation model is then added to the binary star model to 
create the light curve model. 
In this ELCsinus model, the pulsations modulate the average flux and
are not scaled by the binary star light curve; thus they are not boosted 
during the brightening events.
The radial velocities are unchanged. 
The {\sc ELCsinus} model thus consists of a light curve, a radial velocity 
curve for each star, and several derived parameters: L2/L1, and the
log g and $V_{rot} \sin{i}$ for each star. All of these are used to 
compute the goodness-of-fit $\chi^{2}$ statistic and parameter 
uncertainty estimates. 

As shown in Fig.~4, we obtain an excellent fit to the light curve,
matching the amplitude, shape, and phase of the brightening events.
The brightening is due to a combination of mutual irradiation and 
tidal distortion during periastron passage of the pair of A stars
on their highly eccentric, nearly face-on orbit.
The system parameters of KOI-54 are reported in Table~2. 
The uncertainties listed are the formal uncertainties and should be
treated as lower limits, as there are many potential sources of 
systematic errors as discussed in \S5.5. 
As an example, the mass and radius of Star 1 from the  {\sc ELCsinus} 
fit yields a log g that is 1.6 $\sigma$ larger than the spectroscopically
measured log g (even though any deviation from the observed log g
incurs a $\chi^{2}$ penalty). Although it is very difficult to 
assess the size of systematic uncertainties, a doubling of the 
formal uncertainties seems reasonable, especially for the 
inclination, masses and radii of the stars.

Note that the star designated as Star 1 is the {\em less} massive star.
Also notice that the pulsations continue to occur during the brightening
events and exhibit a fixed pattern: there is no discernable phase 
drift. The pulsations are matched well with our simple 
modification to {\sc ELC}, indicating that the pulsations are not 
altered by the irradiation and gravitational distortions at periastron.
Indeed, as discussed in the next section, the pulsations are driven 
by events at periastron passage.


\section{Periastron-Pumped Pulsations}
To investigate the pulsations, the best-fit binary star model was
subtracted from the light curve. This leaves the stellar pulsations,
random noise, systematic calibration noise, and any slight mis-matches 
between the model and actual binary star light curve. The 15-sines used 
in {\sc ELCsinus} are not removed, only the binary star component of the 
model is removed.

The beat pattern seen in the light curve indicates that two closely
spaced pulsations are present in the light curve.
This is confirmed by the power spectrum of the pulsation-only light 
curve -- see Fig.~5.
We computed the power spectrum using the {\sc Period04} software package 
(Lenz \& Breger 2005) which uses an iterative least-squares 
fit to all detected sinusoidal terms simultaneously.
Including the 2 dominant peaks in the power spectrum, there 
are 13 clean peaks with amplitude greater than 12 $\mu$mag,
corresponding to a signal-to-noise ratio $\gtsimeq$20. We list 
the 30 strongest pulsations in Table~3. 
Frequencies less than 2~$\mu$Hz (longer than $\sim$6~d) are not included 
in the Table as these long timescales suffer contamination from 
imperfect detrending, but we did include 5 low-frequency terms when 
computing the power spectrum.
The largest spike, F1, is at a frequency of 2.15286 d$^{-1}$ 
(24.9174 $\mu$Hz) and the next largest, F2, at 2.17680 d$^{-1}$
(25.1944 $\mu$Hz), or roughly 
11.15 and 11.03 hours.
These pulsations are perfectly sinusoidal; there are no
harmonics and no evidence of any modulation in frequency 
(caused, for example, by Doppler shifts due to orbital motion).
We cannot tell which star is pulsating, or if both are, and if so,
which pulsations originate on which star. The pulsation phases are equal 
very near (but not exactly at) the time of periastron, thus the beating 
envelope amplitude is largest near periastron and  lowest at orbital 
phase 0.5. 
These two pulsations are also the fastest in the light curve; there 
is no significant power at frequencies above F2. 
More precisely, between 3 d$^{-1}$ (35 $\mu$Hz) and the Nyquist 
frequency for the Long Cadence data at 24.5~d$^{-1}$ (283 $\mu$Hz), 
there are no peaks above 3 $\mu$mag. 
The one 
month of Short Cadence data was also examined, after simply omitting the 
single brightening event during this month, and its power spectrum also 
reveals no signal greater than 3 $\mu$mag out to 720 d$^{-1}$ (8.33 mHz), 
showing a complete absence of any p-mode pulsations. Thus neither star 
is a  $\delta$ Sct star to exquisitely high precision, which given 
their early A spectral type, is somewhat surprising.

The separation between the two largest peaks, in period, 
is 41.771~d, very close to the orbital period 41.805~d. 
Because the pulsations are not completely resolved in the power 
spectrum, their frequency difference is consistent with being 
identical to the orbital period. We assert that these are exactly 
the 90th and 91st multiples of the orbital frequency. This claim 
is confirmed by the other pulsations: 23 of the 30
largest pulsations are also very nearly exact integer multiples 
of the orbital frequency --- see Table 3. Like the two dominant 
pulsations, these other pulsations are also very ``pure'', with no 
measurable deviations  from being perfectly sinusoidal.
(The naming of the pulsations {\em roughly} corresponds to the 
relative strengths of the pulsations, so F5 is the 5th largest 
pulsation. However, this is not an exact match because the amplitudes 
depend on the specific tapering method used, and as additional Quarters 
of data were added, some spikes swapped relative heights.)
We searched for additional patterns in the power spectrum other than 
harmonics of the orbital frequency. This led to numerous detections
of frequency spacings within multiplets at a value of 
$\delta f \sim\,0.132\,$d$^{-1}$ ($\sim$1.53~$\mu$Hz),
e.g., between F6 \& F8, and F8 \& F9 -- see Table 3.
Given this strong coupling of the orbital frequency with the pulsation 
frequencies, it is very likely that the pulsations are a result of a 
resonance between the dynamic tides and one or more free low-frequency 
g-modes --- see \citet{Aerts07} and \citet{Willems02}
for a discussion of tidally-driven pulsations in binary stars. 
We will return to discuss this in \S5.3.


\section{Discussion, Concerns, and Speculation}

\subsection{Binary Star Phenomena} 
KOI-54 is a remarkable binary star system.
In the following section we further discuss various properties of the
system, but first we mention that if the {\it Hipparcos} parallax 
is reliable for the binary, the distance is 318 $\pm$ 71 pc. 
At apastron the stars are separated by 0.79 AU, which translates to 
2.5 milliarcseconds -- a potentially resolvable separation.

\subsubsection{Photometric determination of $e$ and $i$}
The {\sc ELCsinus} model produces an excellent fit to the photometry
and tightly constrains the system parameters. 
To our knowledge, this is the first determination of stellar parameters
for a non-eclipsing double-lined spectroscopic binary based on
the brightening during periastron passage.
This is possible because of (i) the high precision of the {\it Kepler}
photometry and (ii) the strong sensitivity of the brightening 
to periastron passage, and hence to the orbital parameters.
Remarkably, we have found that the photometry alone can determine 
the orbital eccentricity as precisely as the radial velocities!
This is a consequence of the dependence of the amplitude of the 
brightening to changes in eccentricity as illustrated in Fig.~6. 
In this set of simulations, closely matched to KOI-54,
the only parameter that varies is the eccentricity. In particular, 
the lower panel shows how one can estimate the eccentricity to 
$\sim$1\% just from its amplitude, if all other parameters were known.
The strong sensitivity to eccentricity is simply a consequence of
the brightening being due to tidal forces and irradiation that scale 
as the separation of the stars cubed and squared, respectively, and 
the separation of the stars at periastron is linear in the 
eccentricity.
(Note: we are assuming simple tidal distortion aligned along the 
line joining the center of masses; in the general asymmetric case
the tidal force is much more sensitive to the separation, going as 
$F \propto r^{-7}$ --- see Hut (1981) for a discussion.)

In addition to the eccentricity, the {\it{Kepler}} photometry is also 
able to constrain the orbital inclination, even though there are no 
eclipses and no double-humped ellipsoidal variations are present. 
But in fact the ellipsoidal variations {\em are} present -- for
such a highly eccentric orbit the humps have shifted from the usual 
photometric quadrature phases, 0.25 and 0.75, to the phase of periastron
(defined here as phase 1.0), and the humps
have merged. The two humps are not equal in height because the 
orbital inclination is not exactly zero and the orbital ellipses are not 
aligned along our line of sight (the argument of periastron is not $\pm$ 
90 degrees);
thus the two phases of maximum {\it visible} ellipsoidal distortion
are not symmetric 
about periastron. In addition, the inclination and argument of periastron 
determine the orientation of the irradiated hemispheres of the stars to 
our line-of-sight, and this creates a small asymmetry and shift in the 
phase of the brightening. For the geometry of KOI-54, the larger the 
inclination, the more the peak would shift to earlier orbital phases, 
and the narrower the peak and the more asymmetric it would become
(brighter post-peak than pre-peak). Such effects allow the inclination 
to be measured.


\subsubsection{Relative Contributions of Irradiation and Distortion}
Both tidal distortion and the irradiation/reflection contribute to the 
brightening, and Fig.~7 shows the separate contributions of each.
For KOI-54 the non-sphericity is the larger contributor, 
with the reflection component roughly 75\% as large.
The dominance of the ellipsoidal effect is not a necessity, but 
rather a consequence of the specific binary system parameters,
and could reverse in another system
(i.e. reflection effect $>$ ellipsoidal effect).
As expected, the ellipsoidal component is more centrally peaked than
the reflection component, as it is more sensitive to the separation of 
the stars.
Note that the reflection component is slightly asymmetric, being 
brighter after periastron than before. This is a consequence
of the inclination of the orbital plane and the orientation of the 
irradiated hemispheres with respect to the observer.


\subsubsection{Fit to the Radial Velocities}
While the {\sc ELCsinus} model provides a good match to the {\it Kepler} 
photometry, the match to the radial velocities is not satisfactory at 
certain phases.
The overall features are reproduced, and the rapid rise and fall of the 
velocities near periastron are well-fit, but the velocities at phases 
other than periastron are underestimated in a systematic way --- see
the residuals in Fig.~3.
When 
we fit the radial velocities without the photometry, we obtain much 
better agreement:
reduced $\chi^2$=1.00 vs.~1.61 for (102-8=94) degrees of freedom
-- the solid and dashed RV curves in Fig.~3 show the difference when
using and not using the photometry.
The maximum velocity difference between the models is 0.3 \kms\ 
and occurs near orbital phase 0.1, 
agreeing with where the deviation between the actual data and the 
{\sc ELCsinus} fit are prominent. The fit is also poor near orbital 
phase 0.8, and it is unclear if between these two phases the fit would 
show the same systematic underestimate, since most of this phase range 
lacks observations.
As a check, we fit the velocities using a well-tested RV-only code and 
obtained the same results. Also note that {\sc ELC} computes an 
intensity-weighted net radial velocity, so that irradiation and tidal
distortion are taken into account, e.g.\ center of light need not be
the center of mass, especially at periastron. Thus the best-fit 
solution using the photometry is not the best-fit solution for the 
radial velocities alone.
To confirm this, we used the best-fit radial velocity parameters to
compute a light curve, and the result was a very poor match to the 
photometry. The origin of the photometric versus radial velocity 
disagreement is not understood, and we can only speculate at this point. 
The poorest fit radial velocities are the ones with the lowest amplitudes, 
and these are the hardest to measure, as the lines from the two stars are 
most blended. This could account for most of the discrepancy. 
The stellar pulsations will affect the line profiles, thus adding 
jitter and broadening to the velocities which could induce a bias.
The pulsations will also affect the measurement of $V_{rot} \sin{i}$, 
which could explain the wide range of $V_{rot} \sin{i}$ we measure, from 
$\sim$4 to 10 \kms. Finally, we have assumed that the stellar rotation
axes are both aligned with the orbital axis. Although not expected
to be misaligned
(because the strong tidal forces will tend to align the spin axes, and 
the measured $V_{rot} \sin{i}$ are the same for both stars),
a misalignment could induce significant jitter into the radial velocities 
and bias in the $V_{rot} \sin{i}$ estimates.

\subsubsection{Tidal evolution}
With its eccentricity of 0.83, KOI-54 is very near the upper edge of 
the envelope of the distribution of eccentricities for its orbital period,
e=0.89 following Mazeh (2008). The ``periastron period'', as introduced 
by Mazeh (2008), is the period of a circular binary having the same 
semi-major axis as the periastron distance, and for KOI-54 
$P_{peri-dist}$ = 2.8~d. This short timescale suggests that the orbit
should have experienced enough tidal evolution to synchronize over its
main sequence lifetime and possibly reduce its eccentricity from an 
even larger value. However, in his seminal work on the weak friction model 
for tidal interaction, Hut (1981, 1982) has shown that the evolution 
of eccentricity can be complex, and depends strongly on the equilibrium 
ratio of orbital and rotational angular momentum, $\alpha$.
In particular, for high values of $\alpha$, the evolution of eccentricity 
need not be monotonically decreasing, as the spin and orbital angular 
momentum can be exchanged back and forth. Furthermore, evolution of the 
stellar interior
and its changing moment of inertia can alter simple orbital evolution
(see Zahn 2008 for a discussion).
While there is over a factor of 10 uncertainty in $\alpha$
due to dependence on the unknown internal mass distribution
(parameterized by the radius of gyration) and the poorly known 
stellar rotation period, the ratio of orbit-to-spin angular 
momentum is very large: $\sim$ hundreds to thousands.
This is well over $\alpha_{crit}$=35.447, and tells us that the
evolution of the orbit can be quite complex, with periods of
increasing as well as decreasing eccentricity (Hut 1982).

Tidal forces can synchronize the stellar spin on a timescale 
2-3 orders of magnitude faster than circularization (e.g. Zahn 2008),
so despite the large eccentricity, we can expect the stellar rotation
to have experienced significant evolution towards synchronicity with 
the orbit. The ratio of pseudosynchronous frequency to the orbital 
frequency is a function of the eccentricity only (Hut 1981) (see eq.~1),
so we can compare this with the observations. 
The pseudosynchronous rotation frequency is
$\Omega_{ps}$=2.48 d$^{-1}$,
translating to a spin period of 2.53~d, $V_{rot}$ = 43--47 \kms, 
and a projected $V_{rot} \sin{i}$=4.1--4.5 \kms. 
This value roughly agrees with the lower range of our observed $V_{rot} 
\sin{i}$, and so a pseudosynchronous spin is roughly consistent with 
the observed e, P, and $V_{rot} \sin{i}$.

In our {\sc ELCsinus} model we did let the stellar spins be free 
parameters; these are constrained mainly by the observed 
$V_{rot} \sin{i}$ but there is a weak dependence on the light curve because 
the oblateness of the stars depends on their rotation, and this 
affects the tidal distortion and reflection. 
In principle
the stellar rotation can be refined through the light curve fit,
and this could be used to check for pseudosynchronous rotation.
But in practice this constraint is too weak to be useful.
The models do have a marginal preference for Star 1 to have a rotation 
that is 3.5 times faster than pseudosynchronous, though 
the uncertainty admits the pseudosynchronous rate (there is a $\chi^2$ 
minima at $\Omega_{1} \sim 1$), and such a rotation rate yields a 
$V_{rot} \sin{i}$ that is 2$\sigma$ high compared to the observation.
If Star 1 is indeed spinning faster than pseudosynchronous, 
it is unclear if the star is still spinning down but has
not yet reached pseudosynchronism, or if the star has exchanged
angular momentum with the orbit and is being spun up from the 
pseudosynchronous rate. The observed slow rotation periods compared 
to normal A stars implies that both stars have experienced significant 
spin-down since birth; perhaps a spin-to-orbit exchange of angular 
momentum has prolonged or even increased the eccentricity of the orbit. 
However, Am and Ap stars 
may be born with a very slow spin, so this hypothetical exchange
of spin and orbit angular momentum cannot be inferred.


\subsection{Exploratory Stellar Evolution Modeling}
We have begun a search of parameter space for coeval stellar 
evolution models that fit the stellar constraints as given in
Table~2.  We are using 
an updated version of the Iben (1963, 1965) stellar evolution code 
that includes the OPAL opacities (Iglesias \& Rogers 1996) 
supplemented by the \citet{Alexander1994}
and D. Alexander (1995, private communication)
low-temperature opacities, and 
the Grevesse \& Noels (1993) solar element abundance mixture.  
As a starting point,
we evolved a M$_{1}$=2.33 \Msun\ and a M$_{2}$=2.40 \Msun\ model with 
solar abundances: X=0.68 (H mass fraction), Y=0.28 (He mass fraction), 
and Z=0.04 (mass fraction of elements heavier than H and He). 
We find that the constraints in Table~2 are difficult to match with
solar-metallicity Z=0.02 models, especially the combination of 
temperature and radius: the radii are too small for the temperatures. 
A metal abundance  higher than Z=0.04 would allow a wider range of 
parameters of the evolution models to match the 
(M$_{1}$,R$_{1}$,T$_{1}$)--(M$_{2}$,R$_{2}$,T$_{2}$) pair, 
and indeed the spectroscopy does indicate enhanced metallicity.
However, for this initial investigation we did not further pursue 
the metallicity as this would also require varying the Y abundance
(beyond the scope of this investigation) and also that these 
stars may be metallic A stars (Am) so the surface abundances 
from spectroscopy may not be representative of the entire star. 

In selecting models, we emphasized the constraints on the {\em ratios} 
of temperature, luminosity, and radius over their actual 
values as the ratios are better determined observationally.
We discounted the spectroscopic log~g measurements that suffer from 
large uncertainties. To be specific, we sought to match the ratios in 
temperature   T$_{2}$/T$_{1}$=1.035  $\pm$0.034, 
luminosity    L$_{2}$/L$_{1}$=1.22   $\pm$0.04, 
and in radius R$_{2}$/R$_{1}$=1.06   $\pm$0.02.
As expected, a family of solutions were found; for example a 
pair of stars with temperatures T$_{1}$, T$_{2}$ of 8498~K, 8667~K
satisfies our matching criteria, as does another pair at 8754~K, 8999~K. 
Looking more closely, in the cooler pair, with an age of 0.354 Gyr, 
the temperature, luminosity, and radius ratios (1.020, 1.26, and 1.08) 
are close to the measured values, but the absolute radii (R$_{2}$=2.52 and 
R$_{1}$=2.35 $\Rsun$) are somewhat larger than our estimated values of 
2.33 and 2.20 $\Rsun$.
In the hotter pair, with an age of 0.263 Gyr, 
the temperature, luminosity, and radius ratios (1.028, 1.25, and 1.06) 
are again close to the measured values, and the absolute radii  
(2.28 and 2.15 $\Rsun$) are closer to the reported values, 
but the temperatures are at the high end of the uncertainty range.  
From this exploratory investigation, the important conclusion we can 
draw is that the 
(M$_{1}$,R$_{1}$,T$_{1}$)--(M$_{2}$,R$_{2}$,T$_{2}$) set we
measure is consistent with stellar evolution models of identical 
age and metallicity stars.


\subsection{Tidal Pulsation Modeling}

By way of brief introduction, linear perturbation theory can be used to 
determine the pulsation frequencies in stars, given their internal 
structure. Stellar pulsations are generally of two classes, the p-modes 
or g-modes depending upon whether pressure or gravity is the local 
restoring force. Being acoustic waves, p-mode characteristics are 
dependent on the local sound speed, and are important in the envelope 
and outer portion of the star. Their motion is primarily vertical and 
they have periods of order hours and less. The g-modes are set up by 
the competition between gravity and buoyancy deep within the star.
Their wave motion is primarily horizontal and they have periods of 
order one day (for non-degenerate stars). The g-modes cannot extend 
through convective zones as the 
convective instability means there is no local restoring force. 
The modes are classified according to three spherical harmonic 
integers: the number of radial nodes ($n$), the angular degree ($\ell$), 
and azimuthal order ($m$). (``Nodes'' are the locations where the 
standing wave pattern, caused by the interference of the pulsations, 
has zero amplitude.) The degree $\ell$ gives the number of nodal 
circles on the sphere, and $|m|$ is the number of nodal lines that 
cross the equator (or equivalently, go through the poles). 
Radial modes do not change the spherical symmetry of the star during the
pulsation cycle and are characterized by $\ell=m=0$. The frequencies of
non-radial modes of degree $\ell$ are split by rotation or magnetic 
fields. Rotation lifts the degeneracy in $m$ in such a way that each mode 
of degree $\ell$ results in $2\ell + 1$ multiplet components in the 
frequency spectrum, with $m = -\ell, \cdots, 0, \cdots, +\ell$.
The classical $\kappa$ (opacity valving) mechanism caused by He 
ionization in the stellar envelopes is predicted to drive p-mode 
pulsations in A to early F-type main sequence stars, while g-mode 
pulsations are not expected in these type stars. For a complete 
reference see \citet{Aerts10}.

The periodic tidal forces experienced by the stars at periastron 
passage are expected to induce stellar oscillations,
including g-modes which otherwise would not be present.
To explore such tidal excitation, we assume one of the stars (M$_1$) 
to rotate rigidly and the other star (M$_2$) to be well described as a 
point source. Whenever the ratio of the external tidal force to the self 
gravity at the star's equator is small, i.e. the dimensionless parameter 
$\varepsilon_{\rm T}\equiv (R_1/a)^3 M_2/M_1 \ll 1 $, 
the tide-generating potential in the eccentric orbit can be expanded 
as a series in terms of spherical harmonics and of the companion's 
mean motion. We follow the method outlined in detail in 
\citet{Willems02}, who have shown that the spherical harmonic 
with degree $\ell$=2 is by far the dominant term in the expansion of 
the tide-generating potential, and only components with 
$m=-2, 0, 2$ occur. The values of these components depend on the orbital 
inclination and eccentricity, as well as on the mean and true anomaly 
of the companion in its relative orbit. Whenever the forcing frequency, 
$\sigma_{\rm T} = k \Omega_{\rm orb} + m \Omega_{\rm rot}$, 
matches the star's free gravity oscillation modes, the tidal action 
exerted by the companion gets enhanced and the particular oscillation 
mode gets resonantly excited with the forcing frequency of the dynamic 
tide (Smeyers et al.\ 1998). Thus, we calculated the $\ell$= 2, m=0 
g-mode eigensolutions for each of these stars using the procedure and 
nonadiabatic non-radial pulsation code of Pesnell (1990) as reported in 
Guzik et al.~(2000). We find that the g-mode spectrum for stars of this 
type is very dense, with period spacings of order 0.03~d. These are 
damped oscillations that would die out without the periodic tidal 
``pumping'', but the damping times are very much longer than the orbital 
period (on the order of 1000 years), so once a mode is excited, it is 
certainly not expected to change amplitude during an orbital period 
-- or the lifetime of the {\it Kepler} Mission.

While still adhering to the constraints on the stars discussed above,
we tweaked the tidal model parameters to find a star whose 
$\ell=2, m=0$ spectrum has one mode which coincides exactly with 
F1 (i.e., 90 times the orbital frequency). This model has a mass of 
2.33\,M$_\odot$, $T_{\rm eff}= 8663\,$K and $R=2.15\,$R$_\odot$ and 
most importantly, its $g_{14}$ mode of $\ell=2, m=0$ has frequency F1, 
proving that standard stellar models exist that can generate a 
resonant tidal excitation. 
(The $g_{14}$ mode has $n$=14 radial nodes in the interior of the 
star, and the $\ell=2, m=0$ is a quadrupole that distorts the star
in the correct directions in response to tidal forcing.)
Even though we did not tune the predictions of the tidal excitation 
theory to fit for KOI-54 in this discovery paper, we expect a frequency 
pattern as in the example treated in \citet{Willems02}, i.e., a 
pattern of resonances at spaced values of the rotational and orbital 
frequency as $k\Omega_{\rm orb} + m\Omega_{\rm rot}$. We applied this 
to all the frequencies detected in KOI-54 and their splitting values, 
which allowed us to deduce that $\Omega_{\rm rot} \sim 0.133$~d$^{-1}$
(1.54 $\mu$Hz) from the occurrence of $m=1$ and $m=2$ tidal 
splittings. This translates into a mean rotation period of 7.5~d 
and an equatorial rotation velocity of some 15 \kms. 
If the orbital and rotational axes are aligned, as assumed in the 
tidal theory, this leads to a prediction of $V_{rot} \sin{i}$ of 
1.5 \kms. This rotation velocity is certainly low compared to the
observed $V_{rot} \sin{i}$, being $\sim 1.7 \sigma$ from our adopted 
value, but we remark that the  $V_{rot} \sin{i}$ measurements were
made in the standard way and did not take into account pulsational 
broadening of the lines. We therefore tentatively accept the size of
the tidal splitting of the lines as compatible with the measured 
$V_{rot} \sin{i}$. Furthermore, it is unlikely that the rotation inside 
the star will be rigid, and so this value of $\Omega_{rot}$ should be 
interpreted as an average value over the depth of the star, effectively
negating concerns that this average rotation velocity is too low compared 
to the observed surface $V_{rot} \sin{i}$. In summary, we have found
a model consistent with the observations that can explain 
{\em all but one} of the 30 largest observed pulsations as either 
harmonics of the orbital frequency or $m=1$ and $m=2$ tidal splittings 
of those harmonics. (F25 remains unaccounted for in this model.)
Despite the very attractive features of this model, it is still very 
exploratory in nature and we do not claim uniqueness of the solution.
Future observations will lead to much better frequency resolution,
allowing a much more refined stellar model. 
We point out that slight deviations of the observed frequencies 
of the splittings from their exact predicted frequencies is 
quite interesting, and probably related to the nonrigidity of the 
stellar rotation. When several years of
{\it Kepler} observations are available, we have the exciting prospect 
of using the observed tidal splittings to allow for an asteroseismic 
determination of the internal rotation law of the star.

\subsection{Pulsations Puzzles}
From their spectroscopic temperatures (T$_{\rm eff} \sim$8500, 8800 K)
both stars are too hot to be $\gamma$ Dor pulsators, and the
absence of any periodicities shorter than 11~h in the
Short Cadence light curve power spectrum confirms this.
Why the stars do not exhibit any $\delta$ Scuti-like pulsations
is not known, though it might be that the stars are just beyond
the blue edge of the classical instability strip.
Presumably no unstable modes (p or g) exist for the stars,
and the stars are intrinsically non-pulsators.
This supposition is supported by the exploratory stellar
models discussed in the previous section.

The pulsations that are observed are those modes for which their free 
g-mode eigenfrequency is close enough to a harmonic of the orbital 
period that they are resonantly driven to measurable amplitudes by the 
periodic tidal (and possibly thermal) forces felt at periastron.
Thus while very many pulsation modes are possible, only those that 
are integer multiples of the orbital frequency and phase-locked with 
the binary orbit are present. 

Some challenges posed by the pulsations in KOI-54 include:
(1) Why are the 90th and 91st harmonics of the orbital frequency the 
strongest pulsations?
(2) Why should one particular mode be strongly excited 
while many adjacent ones are not, i.e., why is orbital harmonic 
91 very strong but 92 completely absent?
(3) What limits the highest frequency excited mode and why is there 
no significant power present at higher harmonics?
(4) Why are some modes tidally split but not others?
(5) If only one of the stars is pulsating, which one is it, and
why isn't the other pulsating?

Interestingly, if we accept the slow stellar rotation inferred from 
the tidal splitting of the pulsations, that would mean the star is 
rotating at only a third of the pseudosynchronous rate.
This would seem unlikely in a simple tidal evolution scenario, but
the very large value of orbit--to--spin angular momentum ratio means 
that exchanges of angular momentum are possible and episodes of 
spin-up and spin down can take place. Or, as mentioned earlier, the 
stars are possibly Ap stars that may have been born rotating very 
slowly. A factor of two improvement in the measured $V_{rot} \sin{i}$ 
would be very valuable; the precision is currently available, but
the accuracy is not.

Given that the stars are similar in mass, radius, and temperature,
have identical ages and metallicities, and experience identical
driving frequencies, a total lack of pulsations from one of the 
stars would be puzzling. And yet the matching of the pulsation
frequencies with a single pulsating star, and the lack of any 
apparent double-set of peaks in the power spectrum, does suggest 
that one star could be responsible for all the pulsations.
In principle, high-spectral resolution, time-resolved spectroscopy 
can determine which of the stars is pulsating (or if both are).
If the spectra are obtained near periastron, the component lines
will be well separated and pulsations in one of the sets of lines
immediately tells us which star is pulsating.
However, the intensity amplitude of the pulsations is very small,
$\sim$ 0.1\%, making this rather challenging. Velocity modulations
causing line profile shape changes are much more likely to be
observable, though again the expected variations will be small.
Further complicating the matter is the need to cover at least one 
full oscillation cycle, thus requiring $>$12 hours, and 
perhaps a multi-longitude campaign. Fortunately, the star is bright 
(Kp=8.380) and the integrations can be long and still temporally
resolve the pulsations. 
The reward for such efforts is high: 
given that the two stars are very similar, if only one star is pulsating, 
it means that the excitation mechanism must be very fine-tuned to
drive the oscillations in one star but not the other. Thus there is
the potential for a precision investigation of internal stellar 
structure, given stars of identical age and metallicity, and nearly 
identical masses.

\subsection{Additional Systematic Concerns}
While we are confident the periastron--pumped pulsating binary star
model is correct, given the complexity of both the physics and
the data calibration, several sources of systematic error are potentially 
present. In this section, we describe several of these, with an assessment
of their effects when possible. These are checks of the robustness of the
overall solution, not detailed investigations that could reveal small
changes at the few $\sigma$ level. Towards that goal, we used a linear 
limb-darkened blackbody approximation instead of stellar atmospheres, 
and locked the stellar rotation to the pseudosynchronous spin frequency.
We begin with a look at the data calibration.

The most significant of the systematic errors in the data arise from the 
difficulty in removing small and abrupt changes in the flux level due 
to changes in pointing, cosmic ray events, safe modes, etc.
While such breaks are usually visible in {\it Kepler} light curves, 
the pulsations in KOI-54 hide these. The residuals of the 
{\sc ELCsinus} fit exhibit small but significant  meanderings and tilts 
on longish timescales (days to weeks) which are almost certainly
due to detrending problems. 
Fig.~4 shows the residuals of the data minus the ELCsinus fit {\em{scaled
by a factor of 5 to be more visible}}. The residuals are not white noise,
indicating correlations due to the detrending, but also due to the fact 
that the model includes only 15 sinusoids while at least 30 are present.
The largest set of correlated residuals 
occurs at the brightening event near day BJD 2,455,105  (see middle left 
panel in Fig.~4). Because the determination of the eccentricity and 
inclination are so sensitive to the amplitude and shape of the 
brightenings, even small tilts or jumps could potentially affect these 
estimates. We were concerned about the inclination in particular:
Normally (e.g.\ for eclipsing binaries) a small change in inclination
does not significantly affect the mass estimates for the stars, as
the slope of the sine function near 90 degrees is flat.
But for KOI-54 the inclination is small so changes in $\sin{i}$ go 
linearly with $i$. Because the masses depend on $\sin^{-3}{i}$ a change
from 5 degrees to 6 degrees results in a decrease of 58\% in mass!
To see if detrending imperfections could affect the accuracy of our 
model  parameters (as opposed to precision), we carried out two 
investigations. 
First, we masked out the region with the largest residuals and re-fit. 
Second, we used the power spectrum feature in {\sc ELCsinus} to measure, 
then remove, low-frequency power in the residuals, on timescales of 
10~d and longer. While the $\chi^{2}$ of the models was significantly 
better (by construction), there was no significant change in the system 
parameters.

The determination of the eccentricity depends on the relative flux 
increase of the brightening event. Thus any systematic error in 
normalization will affect the eccentricity. The light curve is in 
fact contaminated by background starlight contained in the same aperture
used for KOI-54, and the relative amplitude of the brightening is thus 
biased low, and hence the derived eccentricity is potentially biased low. 
To check the effect this may have, we subtracted 250 ppm from the light 
curves and re-fit. In addition, we included a small ``third light''
dilution factor into the model and fit for this parameter. In both
cases, there was no significant change in the model parameters.
Furthermore, such a bias would not effect the radial-velocity only
eccentricity estimate, which agrees remarkably well with the photometric 
estimate.

The ``wings'' of the brightening event (i.e.\ the upward curvature 
of the light curve) extend all the way to apastron at phase 0.5; 
there is no phase where the light curve is truly flat.
This is a potential problem in the calibration: we detrended
the light curve assuming it was flat between phases 0.1 and 0.9.
From our binary-star only light curve, the maximum difference in 
relative flux between phases 0.1 through 0.9 is 99 ppm, roughly 1.7 
times the adopted uncertainty per point.
However the light curve is flat to within 25 ppm from phase 0.2 to 0.8,
so the neglect of curvature in the calibration is benign.

In our model, we have neglected several physical effects which we
address below. Systematic errors in the model are potentially more 
serious than those of data calibration, as we have 8 brightening events 
that can help average over any calibration errors. 
For example, 
as is standard practice, we have ignored any consideration of magnetic 
fields. However, for these Am or Ap stars, such neglect at periastron 
passage may not be entirely justified and warrants further investigation.

We have not included relativistic Dopper boosting of the light curves. 
For this nearly face-on binary the boostin effect will be very small, 
and a quick test showed that including the boosting produced a change in 
total (not reduced) $\chi^{2}$ of $\sim$0.1, a completely negligible 
amount. Nevertheless, in general, future models should include this 
effect as a modestly higher orbital inclination would result in a 
measurable difference in the light curve, given {\it Kepler's} 
remarkable precision.

In general, an irradiated star is hotter than an un-irradiated star, 
and so the light that impinges back onto the irradiating star is 
greater than if the irradiated star were isolated. 
Thus the reflection effect should be 
iteratively computed, and this is the standard method championed by 
\citet{Wilson90}. However, for KOI-54 this is a very small effect 
as the total brightening is less than 1\%, and tests with iteration 
showed almost no difference compared to a single calculation assuming a 
point source irradiator (which is exact for perfectly spherical stars). 
Given the large computational cost of full geometry tile-by-tile iteration, 
and the negligible benefit, we used the point-source irradiation 
approximation. Our models have a maximum non-sphericity of 0.7\%
at periastron (ratio of minimum to maximum radii: pole radius to 
L1-direction radius), justifying the spherical approximation for
irradiation. And despite the relatively close periastron passage of
0.065 AU ($\sim$6.4 stellar radii), the stars only fill $\ltsimeq$ 30\% 
of their Roche lobe radii.

The most significant of the limitations of our model is the use of the
Roche approximation. While quite successful at matching the observations,
there is a systematic bias inherent in the use of Roche potentials in 
that the tidal elongations are always pointing along the axis 
connecting the centers of mass. In other words, the stars are treated
as perfect, viscousless fluids that can instantaneously adjust their 
shapes to the gravitational and spin potentials.\footnote{The Roche
potential approximation is similar to the simplest equilibrium tide 
models in this regard.} But real stars will have tidal bulges that
do not instantly re-adjust to the changing external potential, and
are also ``dragged'' by the rotation of the star. Thus the 
elongation will not be aligned with the center of masses except 
when the stars are in synchronous rotation in a circular orbit. 
The irradiation/reflection would also be affected, simply because 
of the different geometry.
Furthermore, the Roche potential treats the mass distribution as
a point-mass at the center of the star; this is equivalent to an 
n=5 polytrope or a tidal Love number $k_{2}$=0. Thus for a given 
external force, the Roche potential yields the least possible tidal 
deformation. It then follows that the amplitude of the ellipsoidal 
variation is minimal in the Roche approximation. While it is true that
the mass inside stars is highly centrally concentrated, a more realistic 
mass distribution would produce a larger tide for a given external force. 
So to match a given observed brightening amplitude, a weaker tidal force 
would be required. This could be achieved by either lower masses, or 
much more likely, a very small increase in separation at periastron due 
to a very small decrease in eccentricity.
Hence the brightening in the Roche potential model is expected 
to be very slightly different than a case where a more realistic 
internal mass distribution is used.

Although we have measured very precise system parameters with
our {\sc ELCsinus} model, for the reasons discussed above the 
accuracy is worse than the precision.
This is especially true for the eccentricity and inclination, and
therefore the stellar masses which are proportional to the
cube of the inclination. On the other hand, the photometric-only
eccentricity agrees well with the radial velocity-only eccentricity, 
indicating that the Roche approximation is valid at the level of 
analysis presented in this discovery investigation of KOI-54. 
A better treatment of the system, well beyond the scope of this 
paper, would involve dynamical tide theory --- particularly relevant 
given the observed tidally-driven pulsations.

This deficiency of the Roche model leads to an interesting supposition.
The angle between the instantaneous centers of the stars and the axis 
of the tidal bulge depends on the internal structure of the star and is 
related to the efficiency of tidal dissipation. Measurement of this
lag angle can then be used to constrain the tidal quality factor $Q$.
If one could determine the instantaneous geometry of the system without 
the photometry, then the lag between the brightening and the true time of 
periastron passage could be measured. The geometry could in principle be 
determined though a great host of radial velocities, corrected for the 
distortions caused by the pulsations. Another possible way is related 
to the pulsations themselves; they provide a clock that could be used to 
set a fiducial time against which to measure the time of brightening. 
Interestingly, the time when the two largest modes (F1 and F2) are in 
phase is not the time of periastron as defined by the photometry, but 
0.39$\pm$0.07~d (=9.4~h)
earlier. This 
can actually be seen in the light curves themselves, as the asymmetric 
shoulders of the brightening and dip at the very pinnacle -- see 
Figs.~1 and 4.

\subsection{KOI-54 in Context}
While tidally-excited oscillations have been studied in quite 
some detail from a theoretical point of view (e.g., see \citep{Aerts10}
for a discussion and references), observational evidence of 
their existence is scarce. Prior to KOI-54, only 3 such systems
were known to exist. The rarity of these stars
is understandable if one realizes that the forcing frequencies 
from the dynamical tides must come very close to the free eigenmode 
frequencies of one of the stars in order to resonantly excite
the oscillation, e.g., \citet{Willems03}. 
This naturally leads to 
the excitation of particular gravity modes whose frequencies are 
{\em exact} multiples of the orbital frequency.

Two of the previously known cases, HD\,177863 (De Cat et al.\ 2000) and 
HD\,209295 (Handler et al.\ 2002) were discovered from ground-based 
photometry and spectroscopy and are single-line spectroscopic binaries.
HD\,177863 is a slowly pulsating B star with $e=0.60, 
P_{\rm orb}=11.9\,$d with one detected gravity mode whose frequency 
is 10 times the orbital frequency.
HD\,209295 is a hybrid $\gamma\,$Dor/$\delta\,$Sct 
star in a binary with $e=0.35, P_{\rm orb}=3.1\,$d with five g-modes 
having frequencies which are exact multiples of the orbital frequency, 
besides a free $\delta\,$Sct mode. 
A much more interesting case is the eclipsing double-lined spectroscopic
B-type binary HD\,174884 discovered with the {\it CoRoT} mission
(Maceroni et al.\ 2009). This system has e=0.29, P$_{\rm orb}=3.66\,$d 
and has pulsations with exact multiples of 2, 3, 4, 8, and 13 times the 
orbital frequency and very tightly constrained physical parameters. 
While it is unclear if the lowest-order multiples are due to imperfect 
prewhitening of the orbital curve (i.e., removal of the binary star
contribution to the light curve) one does not expect this for 
harmonics 8 and 13 as all lower-order harmonics should have 
been found as well. Moreover, these two higher frequencies correspond 
exactly to those of free gravity modes of radial order $\sim 10$, which 
points to tidal excitation. 

With its periodic tidal brightening and rich set of pulsations 
(over 30 pulsations at either integer multiples of the orbital frequency 
or at tidally split multiples of the orbital frequency are present), 
KOI-54 now joins this elite set of eccentric binaries that exhibit 
tidally-driven pulsations.

\section{Summary}
Far from being an ordinary A star, the exquisite {\it Kepler} 
observations of KOI-54 have revealed the object to be a fascinating 
binary star system exhibiting a host of interesting phenomena. 
We have successfully matched the light curve and radial velocities
with a model consisting of a pair of A stars on a highly eccentric 
(e=0.83) orbit, seen nearly face-on ($i$=5.5 degrees).
As the stars closely pass each other at periastron, coming within
$\sim$6 stellar radii, the stars tidally distort each other's shape 
and mutually irradiate and heat each other. The combination of the tidal 
ellipsoidal variation and the irradiation/reflection effect creates 
the periodic 0.7\% brightening seen every 41.8 days. 
In addition, the close periastron passage is responsible for exciting 
a rich set of stellar pulsations with at least 30 modes. 
The two largest pulsations, at the 90th and 91st harmonic of the 
orbital frequency, beat against each other producing 
the modulation envelope seen in the light curve. The remaining
pulsations are explained as additional harmonics of the orbital 
frequency or tidally-split harmonics of those frequencies, 
clearly establishing the tidally-driven origin of the pulsations.
Using these frequencies we are able to deduce the average rotation 
period of the pulsating star, 7.5~d. KOI-54 is the only case 
where the tidally-excited oscillations have allowed such a
measurement. 
Future {\it Kepler} observations will allow factors of several 
times higher precision frequency measurements, offering the
potential to map the internal rotation profile of one of the
KOI-54 stellar components.

Despite our success in modeling the system, many puzzles remain.
For example, it is not known which of the stars is pulsating, or if 
both are. Given that the stars are similar in mass and radius, and 
identical in age and metallicity, it is not known why one star would 
pulsate and the other not. Finally, although the pulsations and 
brightening events appear quite dramatic in the light curve, the 
amplitude of the pulsations is in fact considerably less than 0.1\%;
the discovery of the nature of this remarkable star system was made 
possible by the extraordinary precision and duration of the 
{\it Kepler} photometry.

\acknowledgments
{\it Kepler} was selected as the 10th mission of the Discovery 
Program. Funding for this mission is provided by NASA, Science 
Mission Directorate.
The authors acknowledge support from the {\it{Kepler}} Participating 
Scientists Program via NASA grant NNX08AR14G.
CA and WZ received funding from the European Research Council under the
European Community's Seventh Framework Programme (FP7/2007--2013)/ERC 
grant agreement n$^\circ$227224 (PROSPERITY).
Some of the data presented herein were obtained at the W.M.\ Keck 
Observatory, which is operated as a scientific partnership among
the California Institute of Technology, the University of California, 
and the National Aeronautics and Space Administration. The Observatory 
was made possible by the generous financial support of the W.M.\ Keck 
Foundation.
We thank Mr.~Gur Windmiller for general assistance and for a careful
reading of this manuscript.
We especially thank the many members of the {\it Kepler} Team whose
hard work made these observation possible.
Finally, we thank the anonymous referee for a thorough review of this 
paper.

{\it Facilities:} 
\facility{{\it{Kepler}}},
\facility{Keck:I(HIRES)},
\facility{FLWO:1.5m(TRES)},
\facility{HET(HRS)},
\facility{Lick:Shane(Hamilton Echelle Spectrograph)},
\facility{McD:2.7(Coude Spectrograph)},
\facility{NOT(FIES)}


\clearpage
\begin{figure}
\vspace*{3cm}
\includegraphics[scale=0.70,angle=-90]{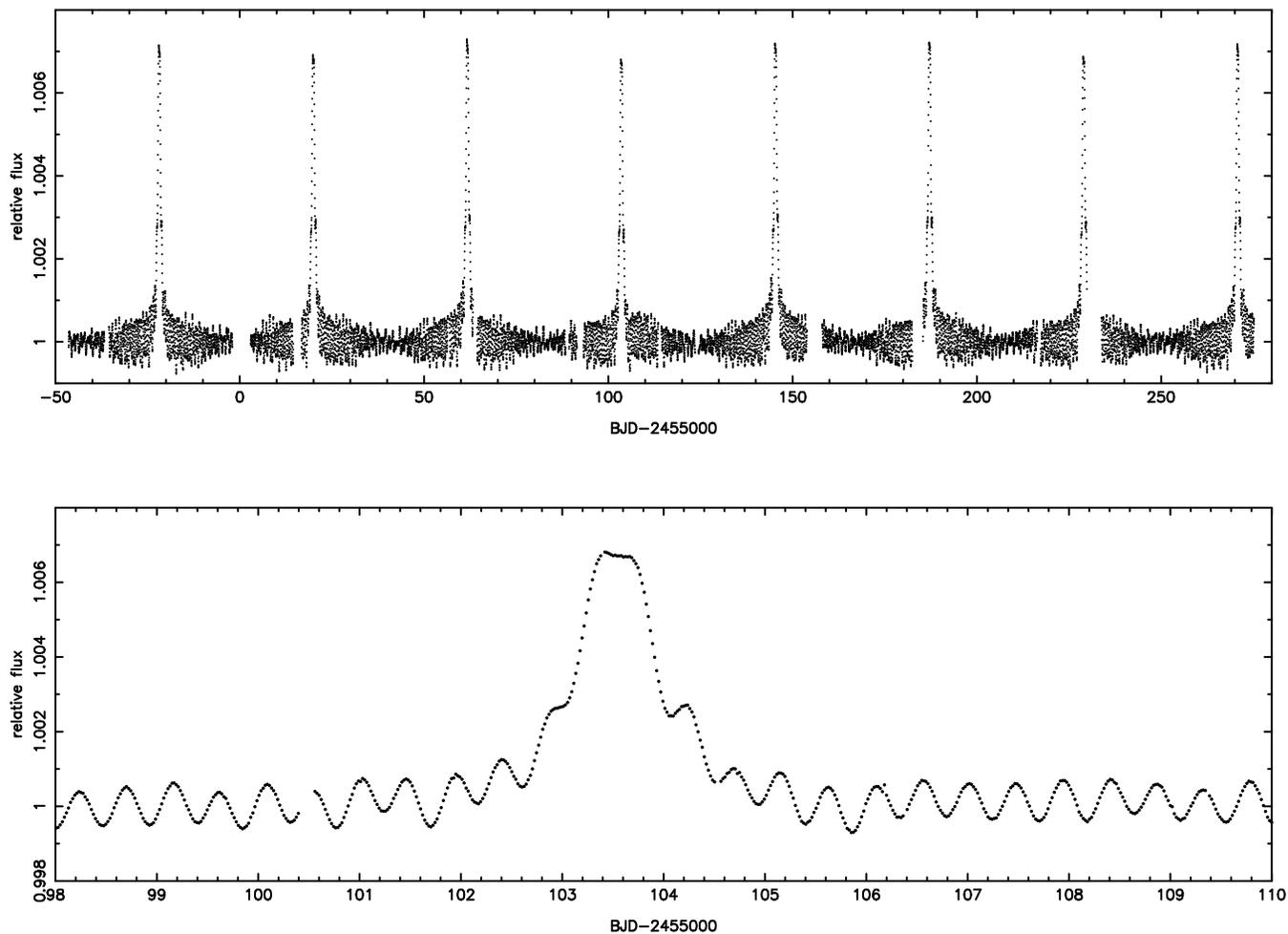}
\vspace*{2cm}
\caption{
{\it{upper}:} The detrended and normalized {\it Kepler} light curve of 
KOI-54.
{\it{lower}:} A detailed view of a brightening event.
\label{fig1}}
\end{figure}
\clearpage
\begin{figure}
\vspace*{3cm}
\includegraphics[scale=0.70,angle=0]{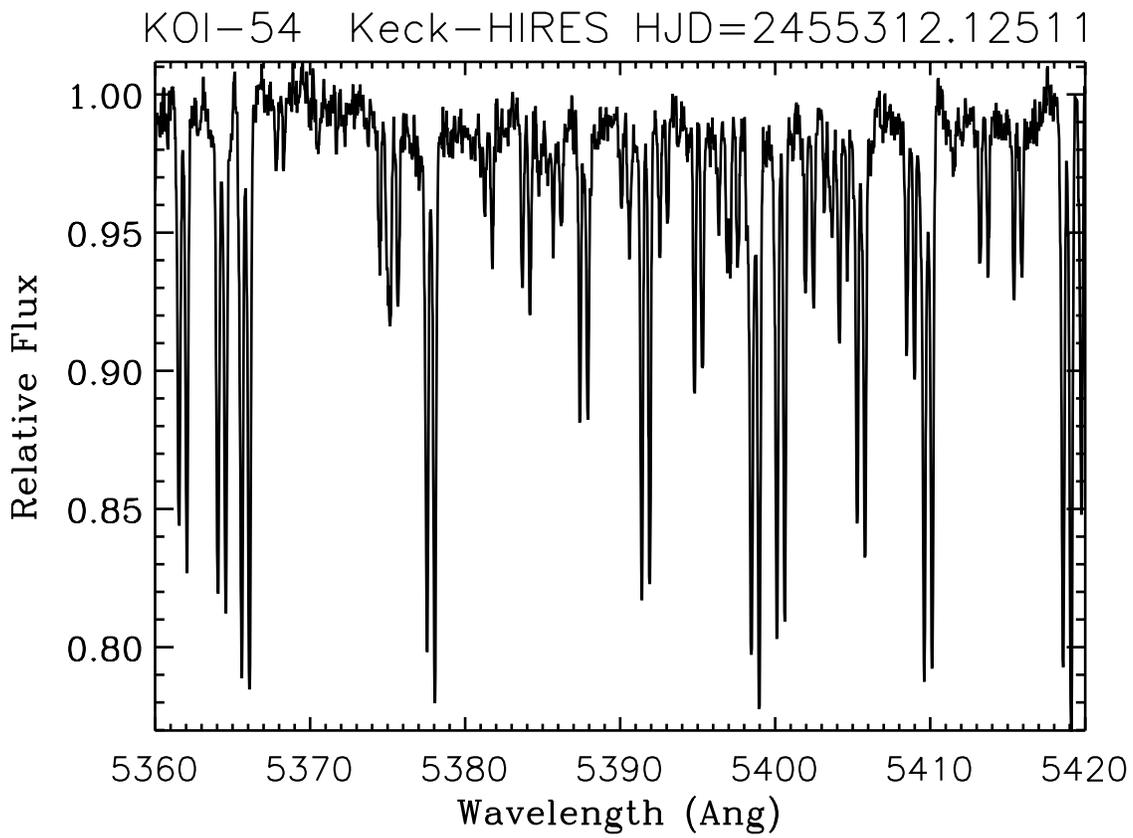}
\vspace*{2cm}
\caption{
Keck HIRES spectrum showing the double-line absorption and
revealing the binary nature of KOI-54.
\label{fig2}}
\end{figure}
\clearpage
\begin{figure}
\vspace*{3cm}
\includegraphics[scale=0.70,angle=0]{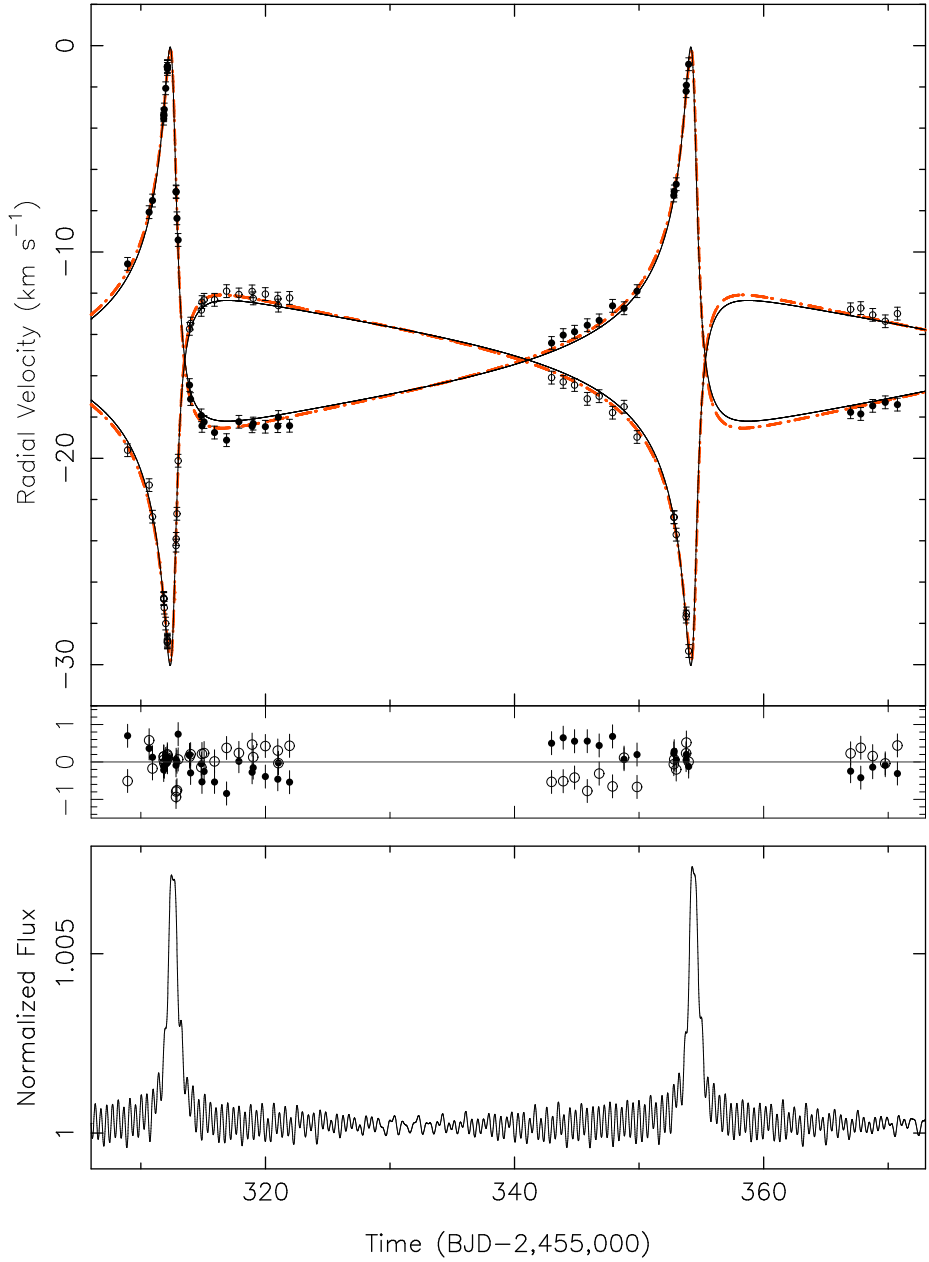}
\vspace*{1cm}
\caption{
{\it{upper:}} Observed radial velocities (RV) and fits: the dashed line
is the fit to the RV data only, the solid line is the best fit model 
that simultaneously fits the RV and the light curve. 
The filled circles denote the RV curve of Star 1.
{\it{middle}:} The residuals of the data minus the best fit.
{\it{lower}:} The {\sc ELCsinus} model light curve showing the predicted 
fluxes at the time of the RV observations.
\label{fig3}}
\end{figure}
\clearpage
\begin{figure}
\vspace*{3cm}
\includegraphics[scale=0.75,angle=-90]{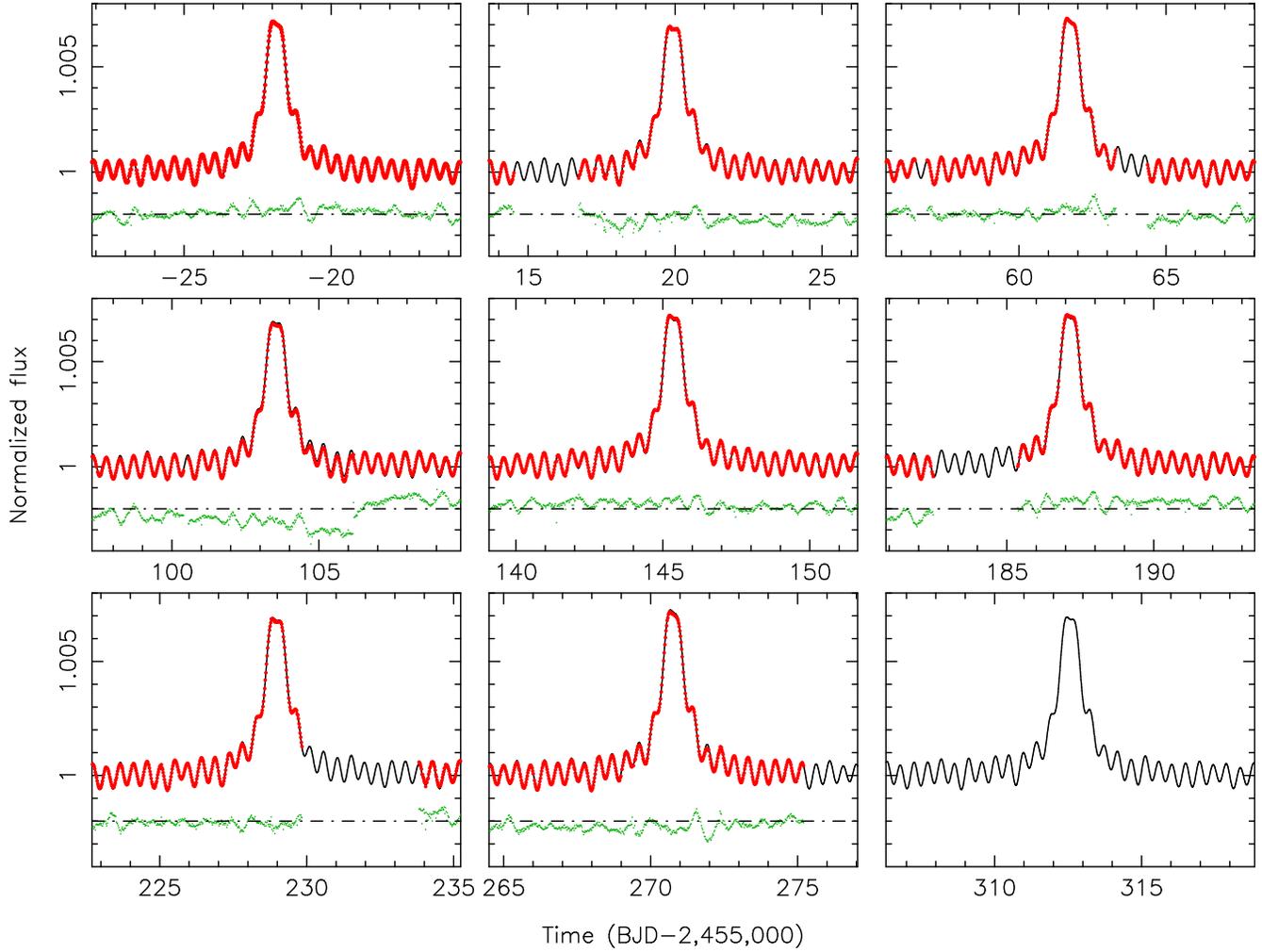}
\vspace*{3cm}
\caption{
Light curves showing individual brightening events with the 
observations plotted as red points, the {\sc ELCsinus} model fit in 
black, and the residuals (offset by +0.998 and scaled by a factor of 5) 
in green. The lower right-hand panel shows just the model.
\label{fig4}}
\end{figure}
\clearpage
\begin{figure}
\vspace*{3cm}
\includegraphics[scale=0.75,angle=-90]{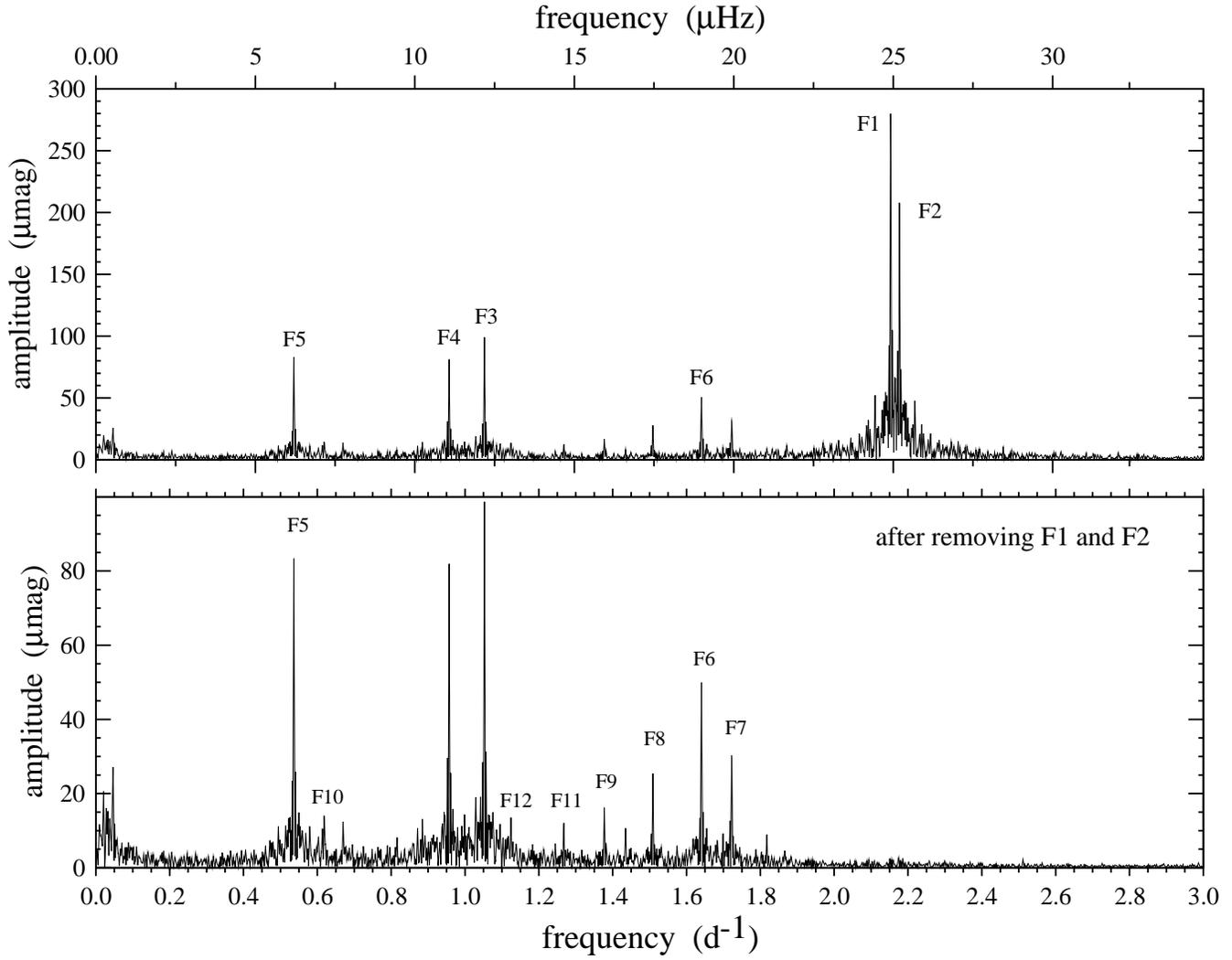}
\vspace*{1cm}
\caption{
Power (amplitude) spectrum of the pulsation-only light curve, 
with the largest
pulsations labeled. The bottom panel shows the power spectrum
after prewhitening by removing the two dominant pulsation F1 and F2.
\label{fig5}}
\end{figure}
\clearpage
\begin{figure}
\vspace*{3cm}
\includegraphics[scale=0.70,angle=-90]{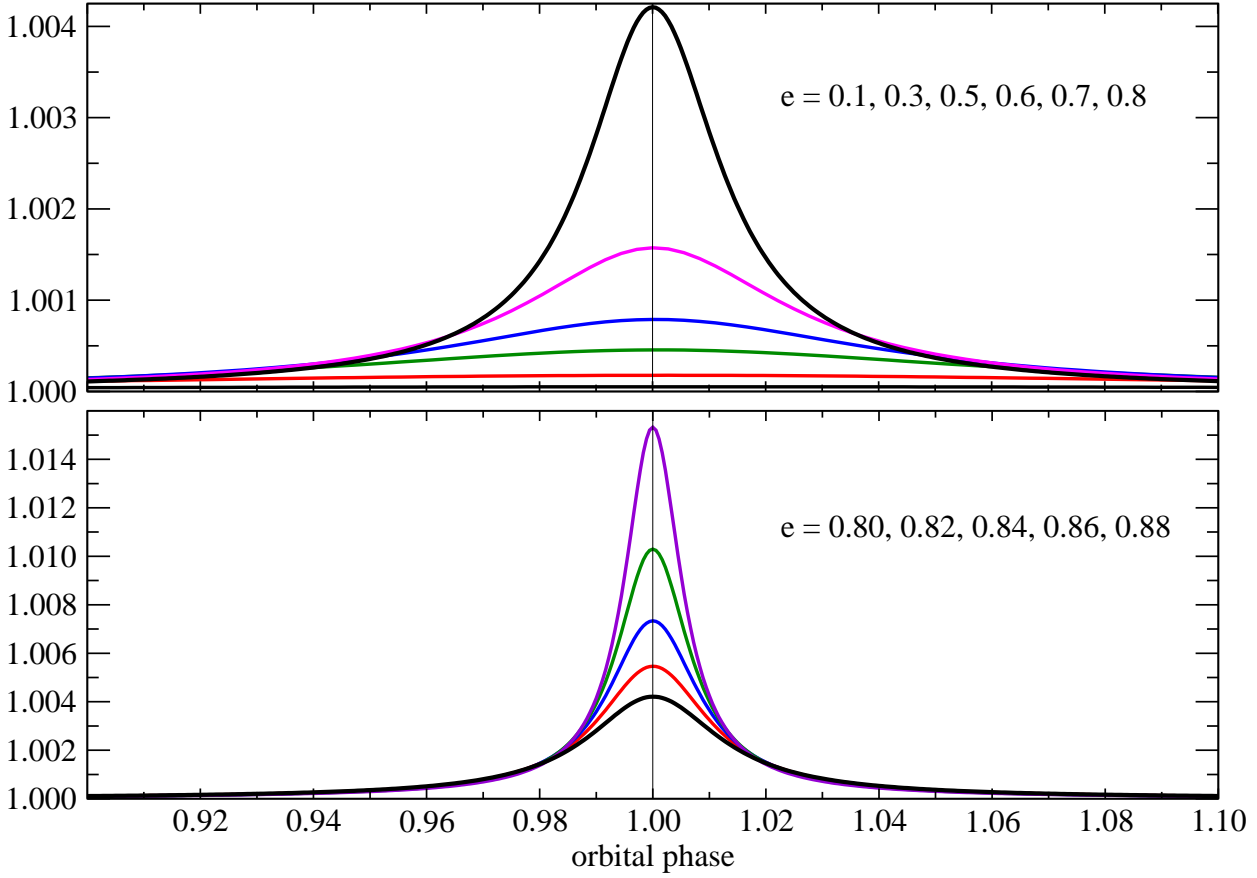}
\vspace*{1cm}
\caption{
Illustrative {\sc ELC} model light curves showing the effect of 
eccentricity on the amplitude of the brightening at periastron.
Eccentricity increases from the lower curve to the upper curve.
The strong sensitivity of the brightening to eccentricity
allows the eccentricity to be determined independently of the radial 
velocities. 
The lower panel shows a tighter range of eccentricities, and by 
inspection, one can see that a brightening amplitude of 0.7\% requires 
an eccentricity near 0.84 for this i=5 degree example.  
\label{fig6}}
\end{figure}
\clearpage
\begin{figure}
\vspace*{3cm}
\includegraphics[scale=0.70,angle=-90]{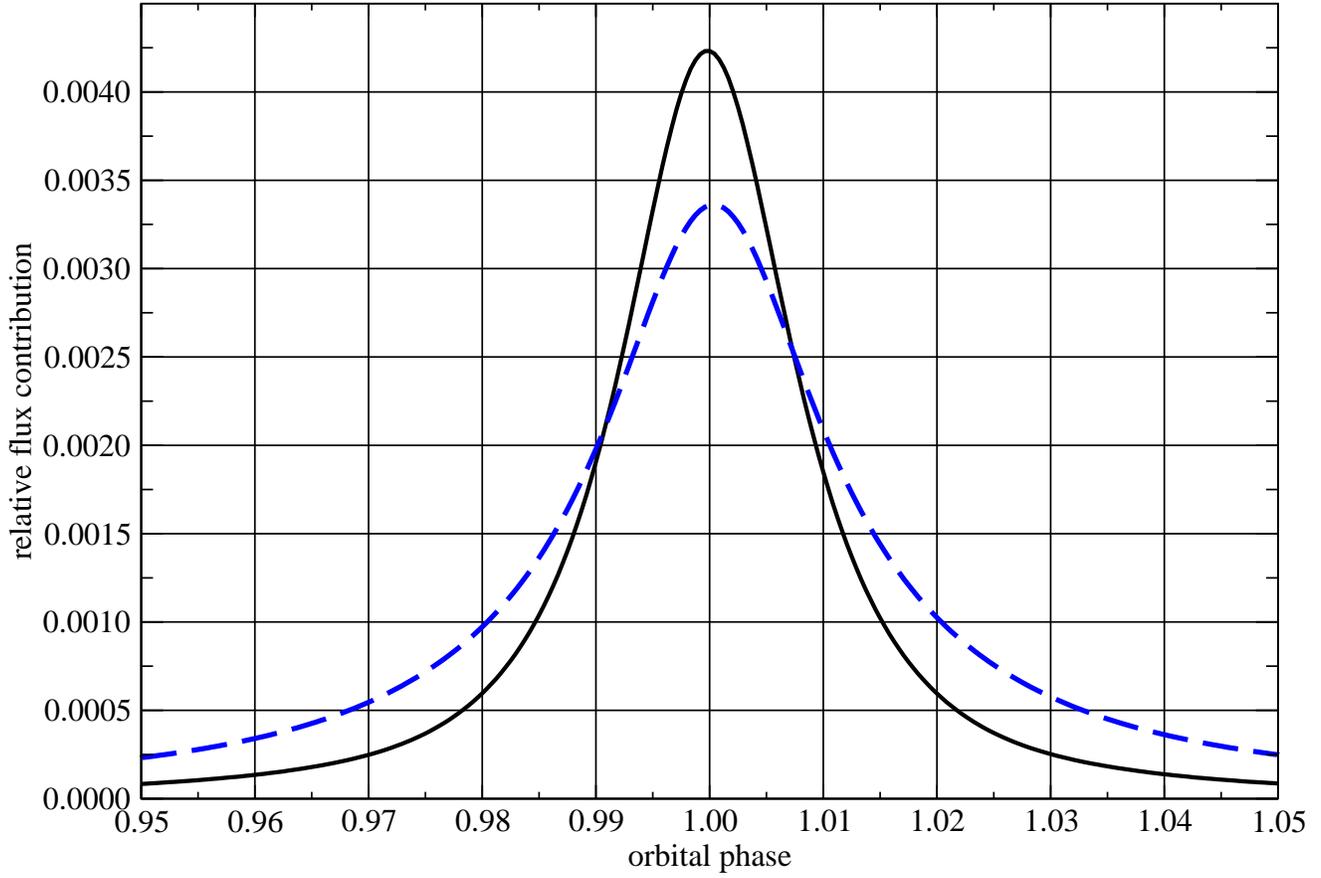}
\vspace*{1cm}
\caption{
Relative contributions to the brightening from the 
tidal/ellipsoidal distortion only (solid curve) and 
irradiation/reflection only (dashed curve).
\label{fig7}}
\end{figure}

\clearpage
\begin{deluxetable}{lrrrl}
\tabletypesize{\scriptsize}
\tablecolumns{5}
\tablecaption{KOI-54 Radial Velocities \label{table1}}
\tablewidth{0pt}
\tablehead{
\colhead{HJD-2400000}   & 
\colhead{RV$_{1}$ }     & 
\colhead{RV$_{2}$}      & 
\colhead{Exp.~Time}     &
\colhead{Facility$^{a}$}
}
\startdata
    55308.9298 &  -10.58 &  -19.61  &     1080 &  TRES  \\
    55310.6587 &   -8.07 &  -21.30  &      900 &  FIES  \\
    55310.9353 &   -7.51 &  -22.83  &      360 &  TRES  \\
    55311.8277 &   -3.54 &  -26.81  &      577 &  MCD   \\
    55311.8353 &   -3.38 &  -26.77  &      547 &  MCD   \\
    55311.8427 &   -3.33 &  -26.82  &      529 &  MCD   \\
    55311.8775 &   -3.10 &  -27.24  &     1800 &  TRES  \\
    55311.9883 &   -2.07 &  -28.01  &     1800 &  TRES  \\
    55312.1251 &   -1.02 &  -28.79  &       72 &  HIRES \\
    55312.1265 &   -1.03 &  -28.93  &       77 &  HIRES \\
    55312.1278 &   -1.13 &  -28.90  &       70 &  HIRES \\
    55312.1292 &   -0.99 &  -28.85  &       72 &  HIRES \\
    55312.8216 &   -7.07 &  -24.23  &      553 &  MCD   \\
    55312.8300 &   -7.10 &  -23.91  &      672 &  MCD   \\
    55312.8899 &   -8.37 &  -22.69  &     1800 &  TRES  \\
    55312.9897 &   -9.42 &  -20.12  &      119 &  HIRES \\
    55313.9043 &  -16.45 &  -13.73  &     1800 &  TRES  \\
    55313.9888 &  -17.13 &  -13.47  &       89 &  HIRES \\
    55314.8513 &  -17.92 &  -12.81  &      543 &  MCD   \\
    55314.9021 &  -18.43 &  -12.43  &      900 &  TRES  \\
    55315.0767 &  -18.24 &  -12.32  &      272 &  HIRES \\
    55315.9130 &  -18.75 &  -12.32  &     1200 &  TRES  \\
    55316.8788 &  -19.12 &  -11.90  &      900 &  TRES  \\
    55317.8700 &  -18.23 &  -12.06  &      900 &  TRES  \\
    55318.9274 &  -18.45 &  -11.91  &     1500 &  TRES  \\
    55319.0179 &  -18.32 &  -12.25  &      144 &  HIRES \\
    55319.9909 &  -18.47 &  -12.04  &     1500 &  TRES  \\
    55320.9937 &  -18.44 &  -12.26  &      840 &  TRES  \\
    55321.0389 &  -18.00 &  -12.60  &       81 &  HIRES \\
    55321.9376 &  -18.42 &  -12.23  &     1680 &  TRES  \\
    55342.9718 &  -14.41 &  -16.09  &     1260 &  TRES  \\
    55343.9095 &  -14.03 &  -16.30  &     1800 &  TRES  \\
    55344.8253 &  -13.87 &  -16.45  &     1260 &  TRES  \\
    55345.8418 &  -13.55 &  -17.12  &     1260 &  TRES  \\
    55346.7917 &  -13.32 &  -16.99  &     1260 &  TRES  \\
    55347.8754 &  -12.61 &  -17.79  &     1800 &  TRES  \\
    55348.7957 &  -12.73 &  -17.50  &     1800 &  TRES  \\
    55349.8360 &  -11.90 &  -18.97  &     1260 &  TRES  \\
    55352.7859 &   -7.26 &  -22.86  &      900 &  HET   \\
    55352.8244 &   -7.07 &  -22.86  &      900 &  LICK  \\
    55352.9850 &   -6.72 &  -23.70  &      900 &  LICK  \\
    55353.7790 &   -2.21 &  -27.67  &      900 &  HET   \\
    55353.7974 &   -1.92 &  -27.52  &      900 &  LICK  \\
    55353.9845 &   -0.90 &  -29.34  &      900 &  LICK  \\
    55366.9826 &  -17.77 &  -12.78  &     1200 &  TRES  \\
    55367.8032 &  -17.85 &  -12.72  &     1800 &  TRES  \\
    55368.7601 &  -17.46 &  -13.05  &     1800 &  TRES  \\
    55369.7711 &  -17.29 &  -13.36  &     1800 &  TRES  \\
    55370.7447 &  -17.40 &  -12.99  &     1800 &  TRES  \\
    55373.7590 &  -17.42 &  -13.17  &     1260 &  TRES  \\
    55376.9083 &  -16.44 &  -14.22  &     1800 &  TRES  \\
\enddata

\tablecomments{
Velocities reported in units of \kms, and exposure times in seconds.
Uncertainties are estimated to be 0.31 \kms.}
\tablenotetext{a}{
Facility Code:\\
TRES:  Tillinghast Reflection Echelle Spectrograph on the F.L.~Whipple 
       Observatory 1.5-m telescope.\\
FIES:  FIbre-fed Echelle Spectrograph on the Nordic Optical Telescope.\\
MCD:   Tull spectrograph on the McDonald Observatory 2.7-m Harlan J.~Smith 
       Telescope.\\
HIRES: HIRES spectrograph on the W.M.~Keck Observatory Keck I telescope.\\
LICK:  Hamilton Echelle Spectrograph on the Lick Observatory Shane 3-m
       Telescope.\\
HET:   HRS spectrograph on the McDonald Observatory Hobby-Eberly 
       Telescope.
}
\end{deluxetable}

\clearpage
\begin{deluxetable}{lrrr}
\tabletypesize{\scriptsize}
\tablecolumns{4}
\tablecaption{ KOI-54 System Parameters \label{table2}}
\tablewidth{0pt}
\tablehead{
\colhead{Parameter}   & 
\colhead{Value}       & 
\colhead{Uncertainty} & 
\colhead{Unit} 
}
\startdata
Star 1 Temperature: T$_{1}$     & 8500  & 200  & K \\
Star 2 Temperature: T$_{2}$     & 8800  & 200  & K \\
log~g$_1$   & 3.8   & 0.2  & (cgs) \\
log~g$_2$   & 4.1   & 0.2  & (cgs) \\
Luminosity ratio: L2/L1 & 1.22 & 0.04 & \\ 
Star 1 V$_{rot} \sin{i} _{1}$ & 7.5  & 4.5 & $\kms$ \\
Star 2 V$_{rot} \sin{i} _{2}$ & 7.5  & 4.5 & $\kms$ \\
Star 1 $[\rm Fe/H]_{1}$  & 0.4  & 0.2  &   \\
Star 2 $[\rm Fe/H]_{2}$  & 0.4  & 0.2  &   \\
\hline
\ \ Fitting RV only: & & & \\
K$_{1}$                           & 9.16     & 0.10    & \kms \\
K$_{2}$                           & 8.85     & 0.10    & \kms \\
Mass ratio, q=M$_{2}$/M$_{1}$     & 1.034    & 0.016   &      \\
Systemic velocity, $\gamma$       & -15.257  & 0.035   & \kms \\
Orbital period, P                 & 41.805   & 0.014   & days \\
Epoch of Periastron, $T_{p}$ & 2455103.5973  & 0.0074  & BJD  \\ 
Orbital eccentricity, e           & 0.8315   & 0.0032  &      \\
Arg. periastron, $\omega$         & 39.46    & 0.51    & degrees \\
\hline
\ \ Fitting RV + light curve: & & & \\
K$_{1}$                           & 9.04     & 0.07   & \kms \\
K$_{2}$                           & 8.82     & 0.09   & \kms \\
Mass ratio, q=M$_{2}$/M$_{1}$     & 1.024    & 0.013   &      \\
Systemic velocity, $\gamma$       & -15.239  & 0.034   & \kms \\
Orbital period, P                 & 41.8050  & 0.0003  & days \\%
Epoch of Periastron, $T_{p}$ & 2455103.5490  & 0.0010  & BJD  \\ 
Orbital eccentricity, e           & 0.8335   & 0.0005  &      \\
Arg. periastron, $\omega$         & 36.70    & 0.90    & degrees \\
Orbital inclination, i            & 5.50     & 0.10    & degrees \\
Semimajor axis, a                 & 0.3956   & 0.008   & AU    \\
Star 1: $\Omega_{1}$              & 3.5      & 2.3     & ---   \\
Star 2: $\Omega_{2}$              & 1.0      & 0.9     & ---   \\
\hline
Star 1 mass: M$_{1}$          & 2.33   & 0.10  & \Msun \\
Star 2 mass: M$_{2}$          & 2.39   & 0.12  & \Msun \\
Star 1 radius: R$_{1}$        & 2.20   & 0.03  & \Rsun \\
Star 2 radius: R$_{2}$        & 2.33   & 0.03  & \Rsun \\
%
\enddata
\tablecomments{
$\Omega$ is defined as the ratio of rotation frequency 
to the pseudosynchronous rotation frequency:
$\Omega \equiv \Omega_{rot} / \Omega_{ps}$
}
\end{deluxetable}

\clearpage
\begin{deluxetable}{lrrrrc}
\tabletypesize{\scriptsize}
\tablecolumns{6}
\tablecaption{ Thirty Largest KOI-54 Pulsations \label{table3}}
\tablewidth{0pt}
\tablehead{
\colhead{ID}            & 
\colhead{frequency}     & 
\colhead{frequency}     & 
\colhead{amplitude}     & 
\colhead{$f/f_{orbit}$} & 
\colhead{nearest}       \\
\colhead{ }             & 
\colhead{(d$^{-1}$) }   & 
\colhead{($\mu$Hz)  }   &
\colhead{($\mu$mag) }   & 
\colhead{ }             &
\colhead{ harmonic }
}
\startdata
F1  & 2.1529 & 24.917 &  297.7 &  90.00  &  90   \\
F2  & 2.1768 & 25.195 &  229.4 &  91.00  &  91   \\
F3  & 1.0525 & 12.182 &   97.2 &  44.00  &  44   \\
F4  & 0.9568 & 11.074 &   82.9 &  40.00  &  40   \\
F5  & 0.5363 &  6.207 &   82.9 &  22.42  &  ---  \\
F6  & 1.6405 & 18.988 &   49.3 &  68.58  &  ---  \\
F7  & 1.7222 & 19.933 &   30.2 &  72.00  &  72   \\
F8  & 1.5087 & 17.462 &   17.3 &  63.07  &  63   \\
F9  & 1.3773 & 15.941 &   15.9 &  57.58  &  ---  \\
F10 & 0.6697 &  7.751 &   14.6 &  28.00  &  28   \\
F11 & 1.2678 & 14.673 &   13.6 &  53.00  &  53   \\
F12 & 1.1241 & 13.011 &   13.4 &  46.99  &  47   \\
F13 & 0.9329 & 10.798 &   12.5 &  39.00  &  39   \\
F14 & 1.4349 & 16.608 &   11.6 &  59.99  &  60   \\
F15 & 0.8851 & 10.244 &   11.5 &  37.00  &  37   \\
F16 & 1.6983 & 19.656 &   11.4 &  71.00  &  71   \\
F17 & 0.6183 &  7.156 &   11.1 &  25.85  &  ---  \\
F18 & 1.8178 & 21.039 &    9.8 &  75.99  &  76   \\
F19 & 0.8574 &  9.924 &    9.3 &  35.84  &  ---  \\
F20 & 0.6458 &  7.475 &    9.1 &  27.00  &  27  \\
F21 & 1.0284 & 11.903 &    8.4 &  42.99  &  43   \\
F22 & 1.0765 & 12.460 &    8.3 &  45.01  &  45   \\
F23 & 1.5092 & 17.467 &    8.1 &  63.09  &  63  \\
F24 & 0.8610 &  9.965 &    6.9 &  35.99  &  36   \\
F25 & 1.4452 & 16.726 &    6.8 &  60.42  &  ---  \\
F26 & 1.2439 & 14.397 &    6.4 &  52.00  &  52   \\
F27 & 1.0078 & 11.664 &    6.3 &  42.13  &  ---  \\
F28 & 0.7894 &  9.137 &    5.9 &  33.00  &  33   \\
F29 & 0.6937 &  8.028 &    5.8 &  29.00  &  29   \\
F30 & 1.1483 & 13.290 &    5.7 &  48.00  &  48   \\
\enddata
\tablecomments{
Uncertainty in frequencies is $\sim$0.0001 d$^{-1}$ or 
0.001 $\mu$Hz.
Formal uncertainty in amplitudes is 0.3 $\mu$mag.
Orbital frequency $f_{orbit}$ was found via least-squares fit
to best match the harmonics: 
$f_{orbit}$ = 0.0239205 d$^{-1}$ = 0.276857 $\mu$Hz.
}
\end{deluxetable}

\end{document}